\documentclass[%
 reprint,
 amsmath,amssymb,
 aps,
floatfix,
]{revtex4-2}

\usepackage{graphicx}
\usepackage{dcolumn}
\usepackage{bm}
\usepackage[utf8]{inputenc}

\usepackage{tikz-cd}

\begin{document}

\preprint{APS/123-QED}

\title{Study of neutrino-nucleus reactions with CRISP Program (0 $<$ $E_\nu$ $<$ 3  GeV)}

\author{R. Perez}
 \email{rvarona90@gmail.com}
\author{A. Deppman}%
 \email{deppman@usp.br}
\affiliation{%
 Instituto de Física da Universidade de São Paulo-IFUSP \\
 Rua do Matão, Travessa R, 187, São Paulo, 05508-090, Brasil
}%


\author{Evandro Andrade-II}
\affiliation{
 Departamento de Ciências Exatas e Tecnológicas, Universidade Estadual de Santa Cruz, Campus Soane Nazaré de Andrade, Rodovia Jorge Amado, km 16, Bairro Salobrinho, 45662-900, Ilhéus-Bahia, Brasil
}

\author{A.R. Samana, F.G. Velasco}
\affiliation{Universidade Estadual de Santa Cruz - UESC, Rodovia Jorge Amado km 16, Ilhéus, 45662-900, Brasil
}%

\author{Guzmán, F}
\affiliation{Instituto Superior de Tecnologías y Ciencias Aplicadas - InSTEC, Avenida Salvador Allende esquina Luaces, La Habana, 10400, Cuba
}%



\begin{abstract}
The neutrino-nucleus reactions are studied at energies from 0 to 3 GeV, using the CRISP program. To simulate these reactions, CRISP uses the Monte Carlo method through an intranuclear cascade model. Quase-elastic and baryonic resonance formation channels for the neutrino-nucleon interaction are considered. The total and differential particle emission cross sections were obtained, obtaining a good agreement with the values reported by the MiniBooNE experiment. The influence of nuclear effects on the studied reactions, such as fermionic motion, the Pauli blocking mechanism, and the nucleonic separation energy, was shown. It was not possible to simultaneously reproduce the $\nu_\mu + D$ and $\nu_\mu + ^{12}C$ reactions using the same axial mass value. For the charged current quasi-elastic channel, $M_A = 0.95 \ GeV$ for the $\nu_\mu + D$ reaction, and $M_A = 1.35 \ GeV$ for the $\nu_\mu + ^{12}C$ reaction. This can be solved if one considers, in addition to the neutrino-nucleon interaction, the neutrino interaction with a pair of nucleons, just as we demonstrate in the last part of this work.
\end{abstract}

\maketitle


\section{Introduction}

Neutrino appears today as one of the most intriguing elementary particles of the Standard Model\cite{Thomas:2001kw}. Its low mass identified by means of the so-called neutrino oscillation\cite{Barger2013} confers to the particle unique properties. However, its weak interaction with matter\cite{formaggio_ev_2012, llewellyn_smith_neutrino_1972} poses serious challenges to the investigation of the particle's properties. Several experiments were developed to investigate the neutrinos properties, as MiniBooNE
\cite{ray_miniboone_2007}, SciBooNE \cite{Tanaka2008}, MINERvA \cite{gran_minerva_2008}, T2K \cite{Abe2011}, MINOS \cite{MINOS_page}, and NO$\nu$A \cite{Nova_page} and new experiments are being under development, as ANNIE \cite{Back2017}, DUNE \cite{Dune_page}, and Hyper-Kamiokande \cite{DiLodovico2017}.

Neutrinos can interact with the nuclei by coherent \cite{Rein1983, Rein2007, Berger2009} and incoherent \cite{Thomas:2001kw, ravndal_weak_1973, rein_neutrino-excitation_1981, berger_lepton_2007, Yang2009} mechanisms. In the coherent form, the neutrino interacts with the nucleus as a whole, and in the incoherent form, the neutrino interacts with the components of the nucleus separately, that is, with protons and neutrons.

In this work, we investigate the incoherent neutrino-nucleus interaction using a Monte-Carlo approach. The neutrino-nucleon interaction is studied utilizing the (quasi)elastic, baryon resonance production, and deep inelastic scattering (DIS) channels. For each of these channels, the charged current (CC) and neutral current (NC) interactions are considered. The nuclear effects taken into account includes the anti-commutation of the fermionic states, the modifications of the nuclear density during the time evolution of the reaction, the thermalization of the nucleus, the formation and decay of baryonic resonances and the pre-equilibrium emission.

To compute the neutrino-nucleon reactions is necessary to introduce some nucleon form factors \cite{Thomas:2001kw, Leitner2006a}. These form factors are functions to adapt the elementary neutrino-quark to the neutrino-nucleon interaction.  A set of form factors frequently used in the literature are the vector $F_{1, 2}^V$, axial $F_A$, pseudo-scalar $F_P$, and strangeness $F_S$ form factors. Using the Conserved Vector Current Hypothesis (CVC) \cite{feynman_theory_1958, towner_currents_1995}, $F_{1, 2}^V$ can be related to the Sachs form factors \cite{stoler_baryon_1993}, which are well known and studied from the electroweak electron-nucleon interaction \cite{nowakowski_all_2005}. Similarly, by the Partially Conserved Axial Current Hypothesis (PCAC) \cite{towner_currents_1995}, $F_P$ can be related to $F_A$. Thus $F_S$ and $F_A$ are left free and have to be parameterized. These form factors are exclusively dependent on the neutrino-nucleon interaction. 

For a correct determination of the form factors, it is necessary to have neutrino-nucleus measurements in the most exclusive way possible. Ideally, measurements of the Charged Current Quasi Elastic (CCQE) and Neutral Current Elastic (NCE) channels should be made separately. The CCQE channel depends only on $F_A$ and therefore can be used to determine this form factor. On the other hand, the NCE channel depends on both $F_A$ and $F_S$, and with $F_A$ determined from the CCQE channel, it is possible to compute $F_S$.

Due to the low neutrino-matter cross section, there is an extra difficulty in setting up the experiments. Hence to date, we have scarce measurement data from the CCQE and NCE channels for the neutrino-nucleus reactions \cite{Mann1977, Baker1981, ray_miniboone_2007, gran_minerva_2008}. Here is where Monte Carlo simulations become helpful. First of all, it is a great tool for comparing theoretical models with experimental data and making the corresponding parameter fits. Secondly, it serves to obtain theoretical results for reactions where experimental data are not available, which can be extremely important in the preparation of future experiments.

The CRISP model \cite{Deppman2004} is a useful tool to investigate several properties of nuclear reactions. It uses Monte Carlo and Quantum Dynamics methods to provide reliable predictions on several aspects of the reaction process. The model can be divided into three parts: The primary interaction, the intranuclear process, and the residual nucleus decay by spallation or fission.

The primary reaction describes the initial interaction of the incident particle with the proton and the neutron in the vacuum. In the CRISP model we can accurately describe the reactions induced by photons \cite{Deppman2002, Deppman2004, Deppman2006}, electrons \cite{Likhachev2003a, Likhachev2003}, protons \cite{Pereira2008, Andrade-II2012}, light nuclei \cite{Abbasi2020, Varona2018}, and in this work we continue the neutrino induced reaction \cite{Vargas2017}. The model can be used also to study ultraperipheral high-energy collisions and production and decay of strange particles. In this work, most of the developments are done in this part of the model, so we postpone a more detailed description to the next sections.

The second part, the intranuclear cascade, represents one of the most advanced aspects of the model. It included a realistic description of the nuclear dynamics before and after the primary interaction, taking into consideration many of the most important nuclear processes that occur in this step of the reaction. Protons and neutrons are described as Fermi gases contained by the nuclear potential. The one-particle states are calculated according to that potential, and in the ground-state, only the lowest levels are occupied. The Pauli principle is considered strictly. After the primary interaction, the movements of all particles in the compound system are considered, which allows a reliable description of the local modifications of the nuclear density due to the momentum transferred to the nucleons by the intranuclear cascade dynamics. The accurate evaluation of the Pauli exclusion principle at every step of the dynamic confers to the model a unique characteristic that allows the precise reproduction of the dynamical evolution of the system without the need of artificial parameters to regulate the outcomes of the reaction. This aspect makes the CRISP model a trustful method to predict the results of nuclear reactions even where no experiments are available to anchor the theoretical calculations. The intranuclear cascade process, in this model, starts as soon as the primary interaction products are created. The cascade continues until the residual nucleus is completely thermalized. Therefore, pre-equilibrium emissions are completely considered in the standard way. The production of nucleons, mesons, and clusters, like deuterons, are considered in this part.

With thermalization, the residual nucleus starts the decaying process, with the emission of nucleons (spallation process) or fission. This process continues until the excitation energy of the residual nucleus is exhaust. In the case of fission, symmetric and asymmetric fission fragments can be generated. Since in the present work this part of the model is less relevant, we address the interested reader to the references.

With that three-step model, CRISP can give accurate and reliable predictions for the entire reaction process. Because of the small number of adjustable parameters in comparison to other models of this class, it has been used for application in Nuclear Reactor Physics and radioactive beam production.

\section{CRISP model}

The CRISP model is a computational program wrote in C++ with the objective of simulating nuclear reactions. The typical situation to use the CRISP is the following: we have an incident particle with energy T and a target nuclei in rest. To simulate this, the CRISP divides the reaction in three fundamental steps: the primary interaction, the intranuclear cascade, and the  evaporation-fission competition. A more complete description of these phases, for the specific case of neutrino-nucleus reaction, is presented below.

\subsection{Primary interaction}

The primary interaction, or the event generation phase, introduces the initial conditions into the CRISP, puts the incident neutrino into the target nuclei, and runs the first neutrino-nucleon interaction. The initial conditions are an incident neutrino with energy T and a target nucleus at rest in the center of a Cartesian Coordinate System. The neutrino moves in the direction of the nucleus and parallel to the z-axis of the Coordinate System. The initial “x” and “y” coordinates of the neutrino in the nuclear surface are aleatory randomized, using a uniform probabilistic density function in the circle $x^2 + y^2 \leq R^2$, where R is the nuclear radius. The “z” coordinate is calculated as $z = -\sqrt{x^2 + y^2}$.

The target nucleus consists of two Fermi gases, protons and neutrons, subjected to a square nuclear potential of depth $V_0 = E_f + B$, where $E_f$ is the Fermi energy and $B = 8$ MeV is the separation energy. The target has a layer structure in the momentum space, with a well-determined occupation number of each level. It is helpful to the implementation of the Pauli principle, so it is not allowed any state with a level with more particles than its occupation number. Initially, the target nucleus is in its ground-state, with the nucleons occupying the lowest energy levels.

Kinematically all particles move in a linear and uniform motion into the studied nucleus. When two particles reach their minimal distance, $b_{min}$, then is considered the possibility of interaction between them. For two particles to interact, the following conditions are necessary: 1) The total neutrino-nucleon cross section, $\sigma_{\nu-N_i}$, must be larger than the geometrical cross section, $\pi b_{min}^2$.  2) Even if condition one is met, the Pauli blocking mechanism must allow the final particle configuration.

When any of the above conditions are not met, then the event generator is reset, and this simulation is counted as an attempted cascade. Otherwise, the event generator is stopped, and subsequently, the intranuclear cascade phase is started.

\subsection{Intranuclear Cascade}

When the event generator ends, we have an excited nucleus product of the first neutrino-nucleon interaction. The intranuclear cascade consists of all the possible reactions for the particles inside the nucleus. In this step, similarly to the event generator, all the particles have a linear and uniform motion. Three kinds of interactions are considered: the particle-particle collision, the particle decay, and the particle arrival to the nuclear surface.

The particle-particle interactions are executed in the same way as the ones described in the event generator phase. These processes usually have more than one resulting channel.  For example, two neutrons can react by the elastic scattering or the inelastic delta-particle plus a nucleon formation. The collision probability of two particles depends on their relative position and momentum, the theoretical cross section, and the Pauli exclusion principle. In the case of the induced neutrino reactions, the following interactions are considered: $NN \rightarrow NN$, $NN \longleftrightarrow NR$, $mN \rightarrow R $, $\pi NN \rightarrow NN $, $mN \rightarrow mN$; where N represents a nucleon, R a baryonic resonance, and m a meson. The implemented resonances are: $\Delta_{1232}$, $\Delta_{1950}$, $\Delta_{1700}$, $\Delta_{1950}$, $N_{1440}$, $N_{1520}$, $N_{1680}$,  and $N_{1535}$. The mesons are: $\pi$, $\Omega$, $\rho$, and $\phi$.

A particle can decay when its mean lifetime is lower than the cascade duration time. The decaying time is randomized using the following pdf:

\begin{equation}
    f(t)=1-e^{-\lambda t},
\end{equation}
where $\lambda$ is the decay constant. The branching fractions and $\lambda$ values for all decaying particles were taken from \cite{tanabashi_review_2018}.

Finally, when a particle reaches the nuclear surface, if its kinetic energy is higher than the potential well, it is ejected from the nucleus. Otherwise, the particle is reflected and continues its motion inside the nucleus. Coulomb's potential and the tunnel effect are considered for charged particles.

The intranuclear cascade ends when there are no particles with sufficient kinetic energy to escape from the nucleus, and there are no mesons or baryonic resonances inside the nucleus. Up to this stage, the CRISP provides the following information: type, energy, and momentum of the emitted particles; and the mass number $A$, charge $Z$, momentum $P$, angular momentum $L$, and excitation energy $E^*$ of the residual nucleus.

\subsection{Evaporation-fission}

After the intranuclear cascade, no nucleon has enough energy to leave the nucleus. The total excitation energy of the nucleus, $E^*$, is generally higher than the nucleon binding energy. The excitation energy can be redistributed in such a way that particle evaporation is possible. In the CRISP, the remaining nucleus of the cascade is considered to be in thermodynamic equilibrium. Then, the relative emission probabilities of protons, neutrons, and alpha particles are calculated using Weisskopf's statistical model. One can also find the relative probability of fission through the theory of Vandenbosch and Huizenga.

As initial data, we have $A$, $Z$, and $E^*$.  In each iteration, the probability of neutron, proton, alpha emission, and nuclear fission are determined. These probabilities are used to aleatory select the event of the iteration.  The evaporation-fission model runs as long as the excitation energy is higher than the minimum of the neutron binding energy $B_n$ and the fission energy $B_f$. If the fission occurs, then the simulation is stopped, and the calculation of the fission fragments is started. When a particle is emitted, $A$, $Z$, and $E^*$ are updated to pass to the next iteration.

\section{Neutrino-nucleon Interaction Model}

The neutrino-nucleon interaction is described through electroweak interactions in the Standard Model Framework. It can be represented by the following diagram:

\begin{figure}[hbt!]
    \centering
    \includegraphics[scale = 0.35]{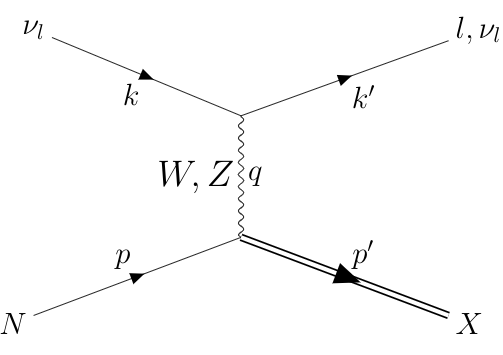}
    \caption{Neutrino-nucleon interaction.}
	\label{fig:diagrama}
\end{figure}

The neutrino $\nu_l$(antineutrino $\bar{\nu}_l$) with momentum $k$ interacts with the nucleon $N$ with momentum $p$ by a boson exchange with a momentum transferred $q$. The resulting particles are the lepton (momentum $k^\prime$) and the baryon or another hadronic system $X$ (momentum $p^\prime$). The contribution of the lepton vertex to the cross section can be calculated in exact form since the neutrino (anti-neutrino) and the corresponding lepton are elementary particles and the coupling of this interaction is well-known from the electroweak formalism.

The hadronic vertex contribution (nucleon--boson--hadronic system) depends on the neutrino interaction with the nucleon being coherent with all constituent quarks or not. Therefore, the type of the neutrino-nucleon interaction is determined by the hadronic vertex, which depends essentially on the neutrino energy. In the following, the implemented neutrino-nucleon channels will be described. In general, the equations depend on the square of the momentum transferred denoted as $Q^2=-q^2$ and given by
\begin{align}
Q^2=2E_\nu E_l - 2 |\vec{k}| |\vec{k^\prime}|\cos\theta - m_l^2    
\end{align}
and 
\begin{align}
    W^2 = M^2 + 2M(E_\nu - E_l) - Q^2
\end{align}
where $E_{\nu}$ is the neutrino energy, $\theta$ is the emission angle of the lepton $l$ in respect to the direction of the neutrino, $M$ is the mass of the nucleon, $m_l$ is the mass of the lepton and $W$ is the invariant mass of the hadronic system produced.

\subsection{Quasi-elastic channel}

In the charged current (CC) the neutrino and the nucleon interact by a boson W exchange, producing a nucleon and the neutrino corresponding lepton. The cross section for this process is given by \cite{leitner_neutrino_2005}:
\begin{align}
\frac{d\sigma^{\nu,\overline{\nu}}}{dQ^2} = \frac{M^2 G_F^2 \cos^2\theta_C}{8 \pi E_\nu^2}\left[ A \mp \frac{s-u}{M^2}B + \frac{(s-u)^2}{M^4}C \right],
\label{CCQE_sigma_dif}
\end{align}
where the negative sign of $B$ is for neutrinos and the positive sign for anti-neutrinos. In the previous equation, $s$ and $u$ are the Mandelstam variables. The A, B, and C parameters depends on the vector $F^V_{1, 2}(Q^2)$, axial $F_A (Q^2)$, and pseudo-scalar $F_P (Q^2)$ form factors.

In the neutral current (NC) the neutrino and the nucleon are elastically scattered after the boson W exchange. In this case, the cross section is \cite{leitner_neutrino_2005}:
\begin{equation}
\frac{d\sigma^{\nu,\overline{\nu}}}{dQ^2} = \frac{M^2 G_F^2}{8 \pi E_\nu^2}\left[ A \mp \frac{s-u}{M^2}B + \frac{(s-u)^2}{M^4}C \right], 
\end{equation}
Now, the $A$, $B$, and $C$ parameters depend on the vector $\tilde{F}^N_{1,2}(Q^2)$, axial $\tilde{F}_A^N(Q^2)$ and strange $F_{1,2,A}^S(Q^2)$ form factors. The superscript $N$ represents the neutron or proton form factor.

In this work, we will discuss the influence of different parameterizations of the $F_A$, $F_1^S$, $F_2^S$, and $F_A^S$ form factors:
\begin{equation}
F_A(Q^2) = \frac{g_A}{(1 + \frac{Q^2}{M_A^2})^2}
\end{equation}
\begin{equation}
F_1^S(Q^2) = -\frac{F_1^S(0)Q^2}{(1+\tau)(1+\frac{Q^2}{M_V^2})^2},
\end{equation}	
\begin{equation}
F_2^S(Q^2) = \frac{F_2^S(0)}{(1+\tau)(1+\frac{Q^2}{M_V^2})^2}, \text{ and}
\end{equation}	
\begin{equation}
F_A^S(Q^2) = \frac{\Delta s}{(1+\frac{Q^2}{M_V^2})^2},
\end{equation}
where $g_A = -1.267$, $M_V = 0.843 \ GeV$, and $\tau = \frac{Q^2}{4M^2}$. The deduction and other form factor expressions for the previous expressions can be found in \cite{Thomas:2001kw,leitner_neutrino_2005,garvey_determination_1993}.

\subsection{Barionic Resonance Formation}
In the CC, the neutrino and the nucleon interact by boson W exchange, producing a baryonic resonance and the neutrino corresponding lepton. In the NC, the neutrino is scattered with the nucleon by the boson W, producing a baryonic resonance.

\subsubsection{$\Delta$-Resonance}

In this case, a different expression is needed for the hadronic current (regarding the quasi-elastic case) and thus a different relation between the form factors and the hadronic tensor. In fact, for the $\Delta^{++}$, one has \cite{lalakulich_resonance_2005}

\begin{widetext}
\begin{eqnarray}
	\frac{d\sigma^2}{dQ^2 dW} = \frac{G_F^2 }{4 \pi} \cos^2 \theta_C \frac{W}{M E_\nu^2}\{ W_1(Q^2 + m_\mu^2) + \frac{W_2}{M^2} [2(k \cdot p)(k^\prime \cdot p) - \frac{1}{2} M^2 (Q^2 + m_\mu^2)] \nonumber \\ -\frac{W_3}{M^2}[Q^2k \cdot p - \frac{1}{2} q \cdot p(Q^2 + m_\mu^2)] + \frac{W_4}{M^2} m_\mu^2 \frac{(Q^2 + m_\mu^2)}{2}-2\frac{W_5}{M^2}m_\mu^2 (k \cdot p)\} \hspace{1.0 cm}
	\label{siggma},
\end{eqnarray}
\end{widetext}
where
\begin{equation}
	W_i =  \frac{f_i(Q^2, E_\nu)}{M\pi}\frac{M_R\Gamma_R}{(W^2-M_R^2)^2 + M_R^2\Gamma_R^2},
	\label{WWW}
\end{equation}
and the functions $f_i$ depend on the form factors $C_i^{V, A}$. The parametrization of these form factors and their relation with $f_i$ is based on Ref. \cite{lalakulich_resonance_2005}. $M_R^2$ is the central mass resonance.

To obtain the NC cross section from the CC expression, it is necessary to multiply the transition vector form factors of the CC channel by the factor $(1-2 \sin^2 \theta_W)$ and use the same transition axial form factors. In addition, the emitted muon must be substituted by a neutrino which means take  $m_\mu$ = 0 in Equation (\ref{siggma}) and drop the factor $\cos\theta_C$ making it 1.

\subsubsection{The Rein and Sehgal Formalism}

The Rein and Sehgal formalism allows the cross section calculation of the resonant channel for all resonances of mass $1 < W < 2$ GeV. The cross section is given by \cite{rein_neutrino-excitation_1981}

\begin{equation}
    \frac{d^2\sigma}{dQ^2 dW^2} = \frac{G_F^2\cos^2\theta_C Q^2}{2 \pi^2 M |\vec{q}^2|}\Big( \Sigma_{++} + \Sigma_{--} \Big),
\end{equation}
with
\begin{equation}
    \Sigma_{\lambda \lambda^\prime} = \sum_{i=L, R, S} c_i^\lambda c_i^\lambda \sigma_i^{\lambda \lambda^\prime}
\end{equation}
and
\begin{eqnarray}
    c_L^\lambda&=&\frac{K}{\sqrt{2}}(j^*_x+i j^*_y) \nonumber \\
    c_R^\lambda&=&-\frac{K}{\sqrt{2}}(j^*_x-i j^*_y) \\
    c_S^\lambda&=&K\sqrt{\big|(j_o^*)^2-(j_z^*)^2 \big|} \nonumber.
\end{eqnarray}

The quantities $K$, $\nu^*_{(\lambda)}$, and $Q^*_{(\lambda)}$ result from relating the reference frame of rest of the resonance (RRS) to the reference frame of rest of the initial nucleon and are given by

\begin{align}
    K&=\frac{|\vec{q}|}{E_\nu \sqrt{2Q^2}}, \\
    \nu^*_{(\lambda)}&=\frac{K\sqrt{Q^2}}{c_S^\lambda}j_z^*, \text{ and}\\
    Q^*_{(\lambda)}&=\frac{K\sqrt{Q^2}}{c_S^\lambda}j_0^*,
\end{align}
where, $j_\mu^*$ are the components of the lepton current at the RRS frame.

The partial $\sigma_i^{\lambda \lambda \prime}$ are calculated as:
\begin{widetext}
\begin{equation}
	\sigma_{L, R}^{\lambda\lambda^\prime}(q^2, W) = \frac{\pi}{\kappa}\frac{W}{m_n}\frac{1}{2} \sum_{j_z}\left|\left\langle R, j_z  \left| F_\mp^{\lambda\lambda^\prime} \right| N, j_z \pm 1 \right\rangle \right|^2 \delta(W-W_0)
	\label{2.123}
\end{equation}
and
\begin{equation}
	\sigma_S^{\lambda\lambda^\prime}(q^2, W) = \frac{\pi}{\kappa}\frac{W}{m_n}\left( \frac{Q^2}{-q^2} \right) \frac{m_n^2}{W^2} \frac{1}{2} \sum_{j_z}\left|\left\langle  R, j_z \left| F_0^{\lambda\lambda^\prime} \right| N, j_z \right\rangle \right|^2 \delta(W-W_0).
	\label{2.124}
\end{equation}
\end{widetext}

To obtain the operators $F_{\pm , 0}^{\lambda\lambda^\prime}$, the Feynman--Kislinger--Ravndal (FKR) relativistic model is used \cite{ravndal_relativistic_1971,feynman_current_1971}. In this model 
the baryon is considered a coupled harmonic oscillator of three quarks. The baryon state is calculated 
as a combination of states of spin, isotopic spin and orbital excitation mixed symmetries such that the resulting state is symmetric (color symmetry is not considered). The operators $F_{\pm , 0}^{\lambda\lambda^\prime}$ are obtained depending on the proposed couplings for the vector
current and the vector-axial current of the considered oscillator Hamiltonian. The expressions 
for all the required elements to obtain the cross section are reported in Ref. \cite{rein_neutrino-excitation_1981}.

\subsection{Deep inelastic scattering (DIS)}

The deep inelastic scattering is important for high energies, where the incident neutrino/anti-neutrino can interact at the quark level with the nucleon and create a 
corresponding lepton plus a hadronic system X. 

The DIS is divided in two steps:

\begin{itemize}
    \item Neutrino-quark interaction and hadronic system X formation: This step determines the interaction cross section. It has a high dependence on nucleon structure through the different structure factors. Kinematically, it depends on the 4-vectors of the incident neutrino, $k$, the lepton, $k^\prime$, the nucleon $P$ and the invariant mass $W$ of the 
    hadronic system.
    \item Hadronization: Formation of the constituent hadrons of system X. This step was not 
    studied in this work.
\end{itemize}

In the following, the equations related to the first step of the DIS are described \cite{formaggio_ev_2012}.
\begin{eqnarray}
    \frac{d^2\sigma^{\nu, \bar{\nu}}}{dxdy} &= \frac{G_F^2ME_\nu}{\pi(1+Q^2/M^2_{W,Z})^2} \Bigg\{ \bigg( 1-y-\frac{x^2 y^2 M^2}{Q^2} \bigg)F_2 \nonumber \\
    &+ y^2x F_1 \pm \bigg( y-\frac{y^2}{2} \bigg)x F_3 \Bigg\}.
    \label{2.236}
\end{eqnarray}
In equation above, $\nu=E_\nu-E_l$ is the energy transferred to the exchange boson, where $E_\nu$ is 
the energy of the incident neutrino and $E_l$ is the energy of the resultant lepton. $Q^2=-q^2$, $x=\frac{Q^2}{2M\nu}$ and $y=\frac{\nu}{E}$.

In the term containing $F_3$, the incident neutrino has positive sign and the anti-neutrino has negative sign. For the CC, the mass of the boson $W$ is used and for the NC the mass of the boson $Z$ is used. The structure functions of the nucleon $F_{1,2,3}(x,Q^2)$ are expressed in relation to the quark distribution functions of the nucleon, $q_i(x, Q^2)$, where $q_i= \{ u,\bar{u}, d, \bar{d}...\}$. In this work, the values of the quark distribution functions were taken from \cite{buckley_lhapdf6_2015}.

\section{Results and Discussion}

In the following, the main results obtained for neutrino-nucleus reactions are presented. The 
experimental data were published by the MiniBooNE experiment \cite{ray_miniboone_2007}. CRISP uses the MiniBooNE $\nu_\mu$ and $\bar{\nu}_\mu$ flux predictions ($0<E<3 \ GeV$) \cite{aguilar-arevalo_first_2010, aguilar-arevalo_first_2013} as input data, which is used as the probabilistic density function to generate the incident neutrino energy.

\subsection{Charged current quasi-elastic channel}
The charged current quasi-elastic channel (CCQE) is observed when the neutrino (antineutrino) interacts with a neutron (proton) and produces a negative (positive) muon plus a neutron (proton): $\nu_\mu (\bar{\nu}_\mu) + n (p) = \mu^- (\mu^+) + p (n)$. If $T_ \mu$ is the muon kinetic energy and $\theta_\mu$ its emission angle, then the neutrino energy and the four-momentum transferred to the neutron can be determined by 
\begin{equation}
	E_{\nu}^{QE} = \frac{2m_n^\prime E_\mu - (m_n^{\prime 2}+m_\mu^2-m_p^2)}{2(m_n^\prime-E_\mu + \sqrt{E_\mu^2-m_\mu^2}\cos \theta_\mu)}
	\label{eq.CCQE1}
\end{equation} 
and
\begin{equation}
Q_{QE}^2 = -\mu^2 + 2E_{\nu}^{QE}(E_\mu - \sqrt{E_\mu^2-m_\mu^2}\cos \theta_\mu)
\label{eq.CCQE2}
\end{equation} 
respectively, where $E_\mu = T_\mu + m_\mu$, $m_\mu$ is the muon mass, $m_p$ is the proton mass and $m_n^\prime$ is an effective neutron mass that depends on the carbon bound energy, i.e.,  we have $m_n^\prime = m_n - E_b$, with $E_b = 34 \pm 9 \ MeV$.

\begin{figure}[hbt!]
	\centering
	\includegraphics[scale=0.4]{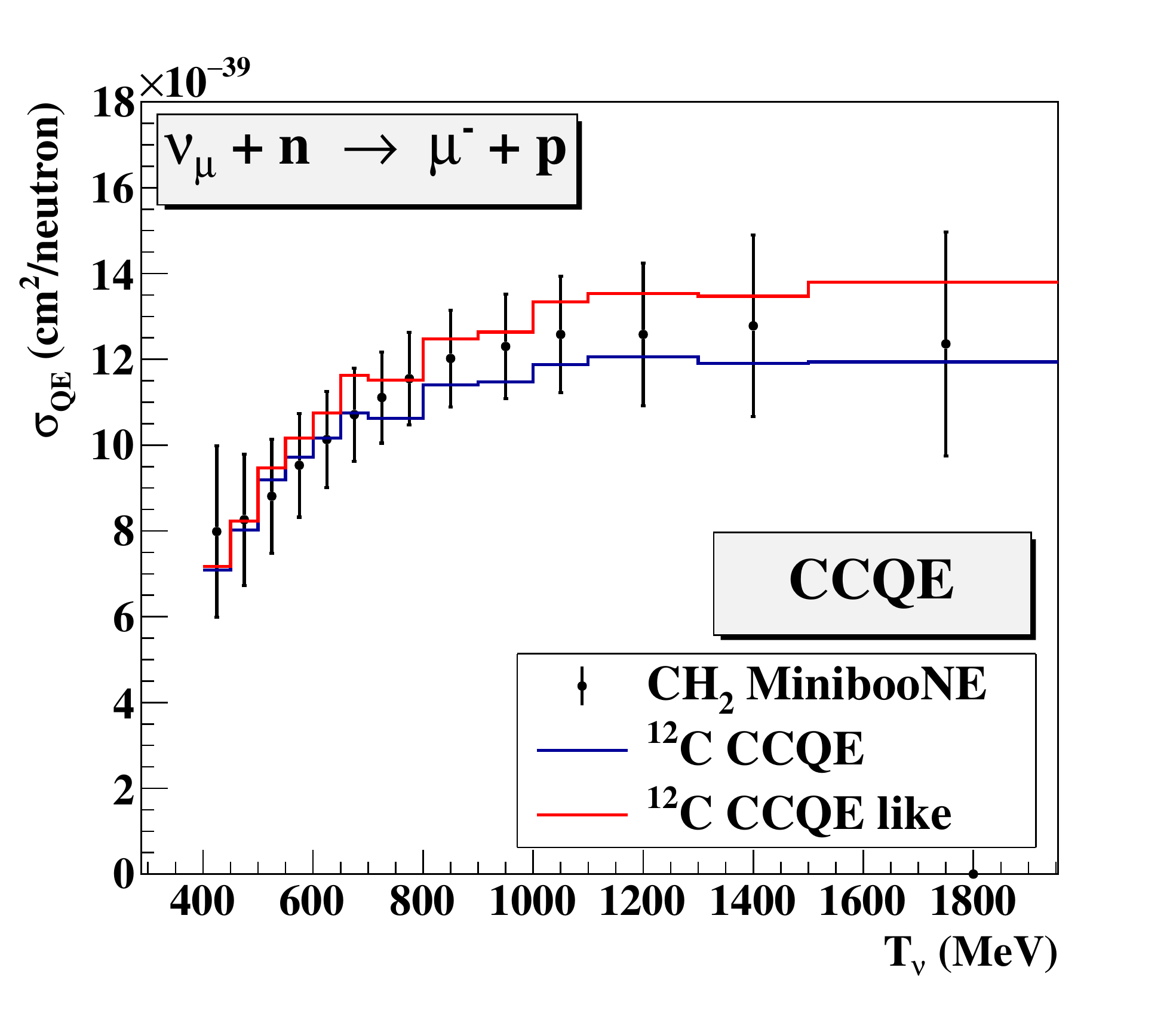}
	\caption{ Total cross section for the reaction $\nu_\mu + n \rightarrow \mu^- + p$. The 
	experimental data correspond to the reaction $\nu_\mu + CH_2$ and were taken from 
	\cite{aguilar-arevalo_first_2010}. The simulations were performed for the reaction $\nu_\mu + ^{12}C$.}
	\label{fig:3}
\end{figure}

In figure \ref{fig:3}, the cross section per neutron for the reaction $\nu_\mu + ^{12}C$ is shown, as a function of the kinetic energy of the incident neutrino. The red line represents the "CCQE like" cross section when a muon and no pions are emitted. The blue line represents the CCQE cross section, after eliminating the CC1$\pi$ contribution from the "CCQE like" cross section.

The CC1$\pi$ contribution is defined when the incident neutrino triggers the following sequence of reactions:

\begin{equation}
\begin{tikzcd}
\nu_\mu (\bar{\nu}_\mu) + n(p) \arrow{r} &
\mu^-(\mu^+) +  N^* \arrow[d, black, shift left = 5.3ex] \\
& \hspace{10.3ex} N^* \arrow{r} & N +\pi,
\end{tikzcd}   
\end{equation}
where $N^*$, $N$, and $\pi$ represent a baryon resonance, a nucleon and a pion respectively.

The difference between the lines in Fig. \ref{fig:3} is caused by the processes of production and absorption of pions in the intranuclear cascade. If the pion formed in a CCpi process is absorbed inside the nucleus,  then the initial CCpi channel can be detected as a CCQE channel. This CCpi background is considered in the experiment when reporting the CCQE cross section \cite{aguilar-arevalo_first_2010}. Figure \ref{fig:3} shows a good agreement between the CRISP simulations and the experimental data.

\begin{figure}[hbt!]
	\centering
	\includegraphics[scale=0.4]{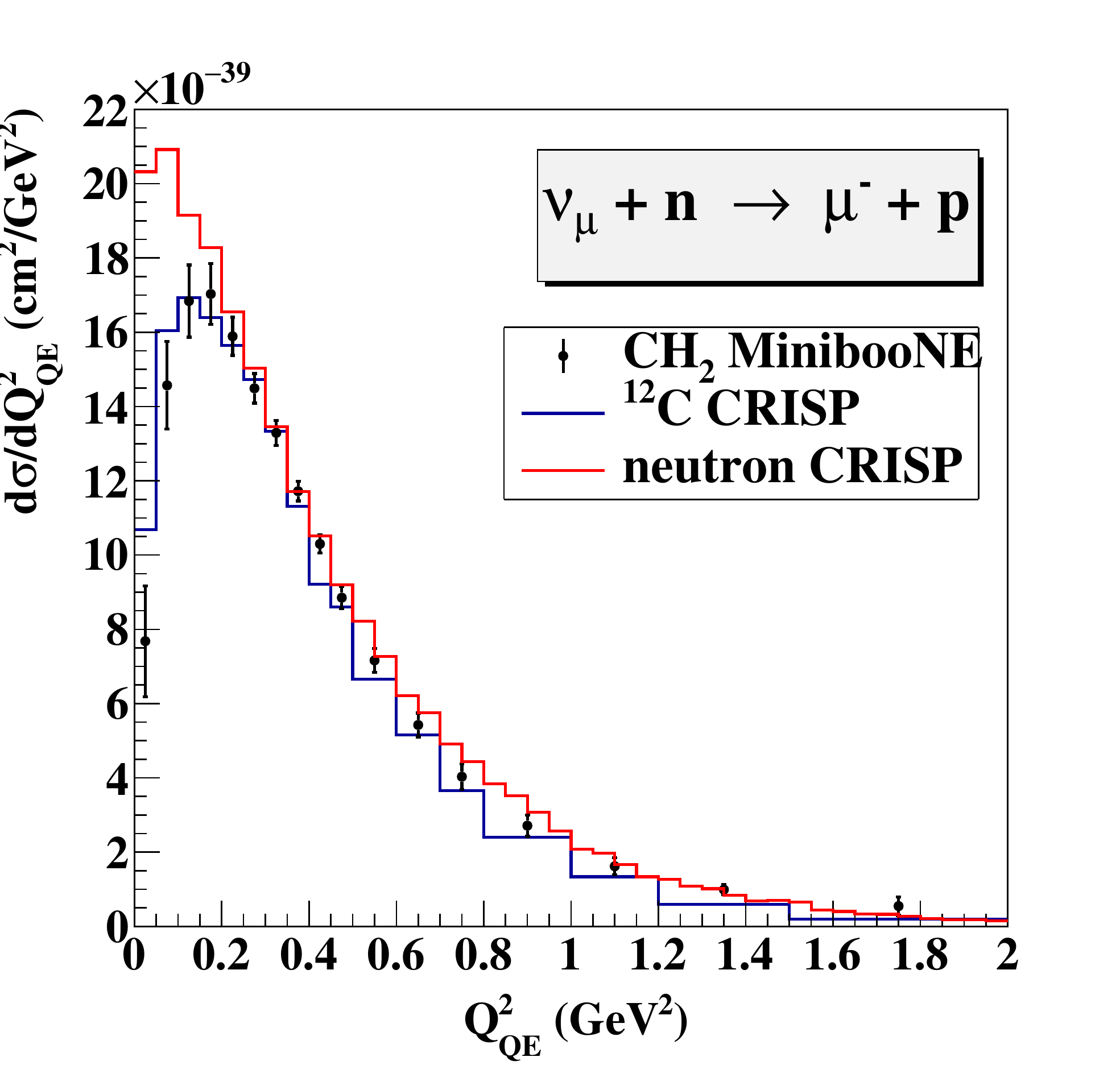}
	\raisebox{0.2\height}{\includegraphics[scale=0.3]{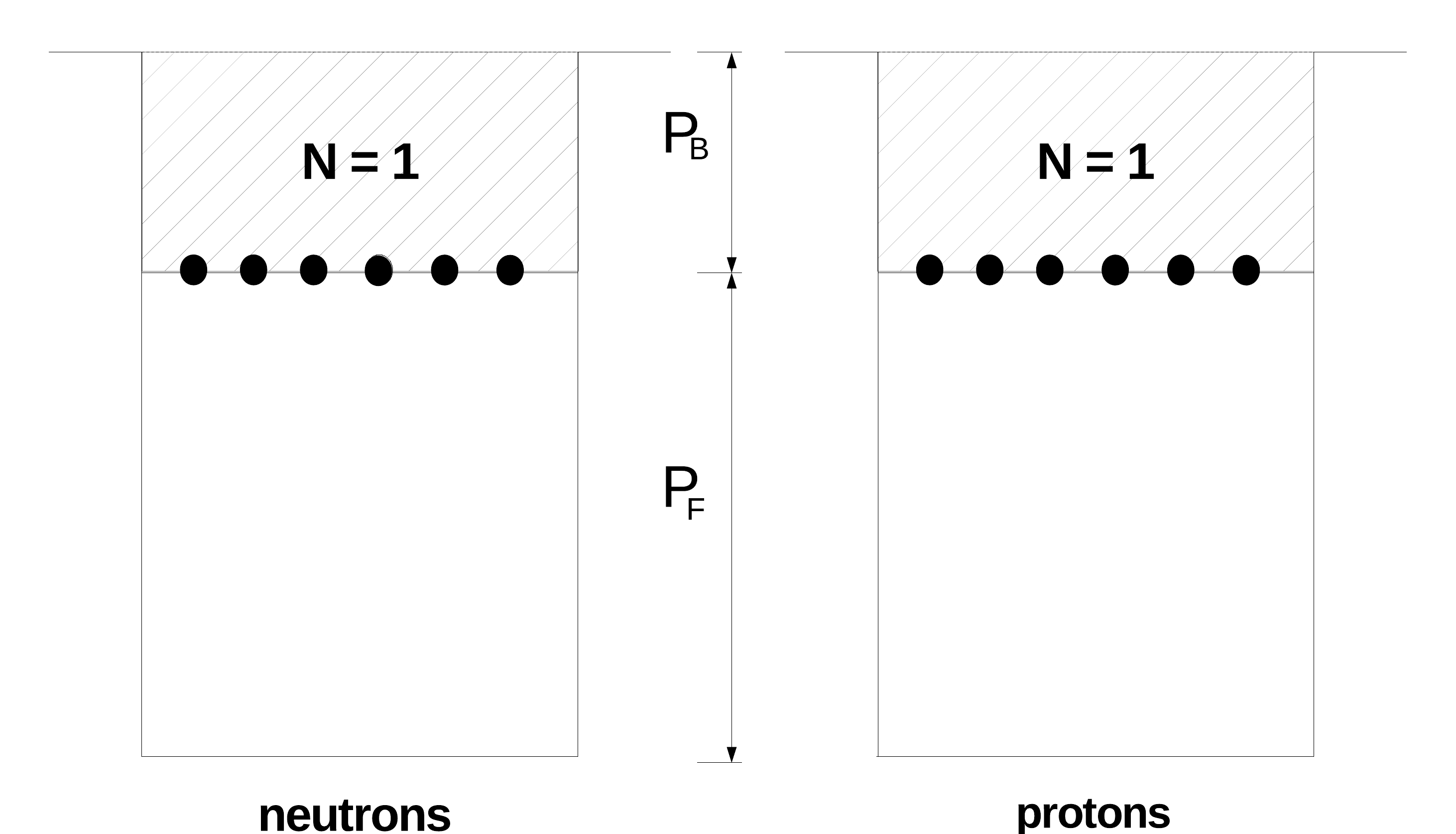}}
	\caption{Top: $d\sigma/dQ^2$ cross section per neutron for the channel $\nu_\mu + n \rightarrow \mu^- + p$ 
	(CCQE) in reactions $\nu_\mu + ^{12}C$ (blue line) and $\nu_\mu + n$ (red line). Bottom: Shell 
	structure for $^{12}C$ used in simulations. $^{12}C$ is formed by one energy level for neutrons 
	(protons). At the fundamental state, the level $N=1$ is completely occupied by 6 neutrons (protons). 
	$P_F$ is the Fermi momentum and $P_B$ is the minimal momentum necessary to force the emission of nucleons.}
	\label{fig:3a}
\end{figure}

In Fig. \ref{fig:3a} (top), the differential cross section $\frac{\partial \sigma}{\partial Q^2}$ 
(blue line) is presented, in which one can observe an overestimation of the cross section for 
$Q^2<0.2$  $GeV^2$. The red line in Fig. \ref{fig:3a} represents the cross section $\nu_\mu + n$, 
for the free neutron. One can see that even by using a simple Fermi Gas nuclear model, it is 
possible to reproduce the nuclear effect ($Q^2<0.2$  $GeV^2$).

The Fig. \ref{fig:3a} (bottom) represents the Fermi Gas model used to build the target nucleus. 
In this model, carbon nucleus is composed of only one energy level completely occupied by 6 nucleons. 
Thus, the Pauli blocking mechanism may occur if the neutrino interacts with any neutron and 
the formed proton falls into the level $N=1$ of the proton well.

\begin{figure}[hbt!]
	\centering
	\includegraphics[scale=0.4]{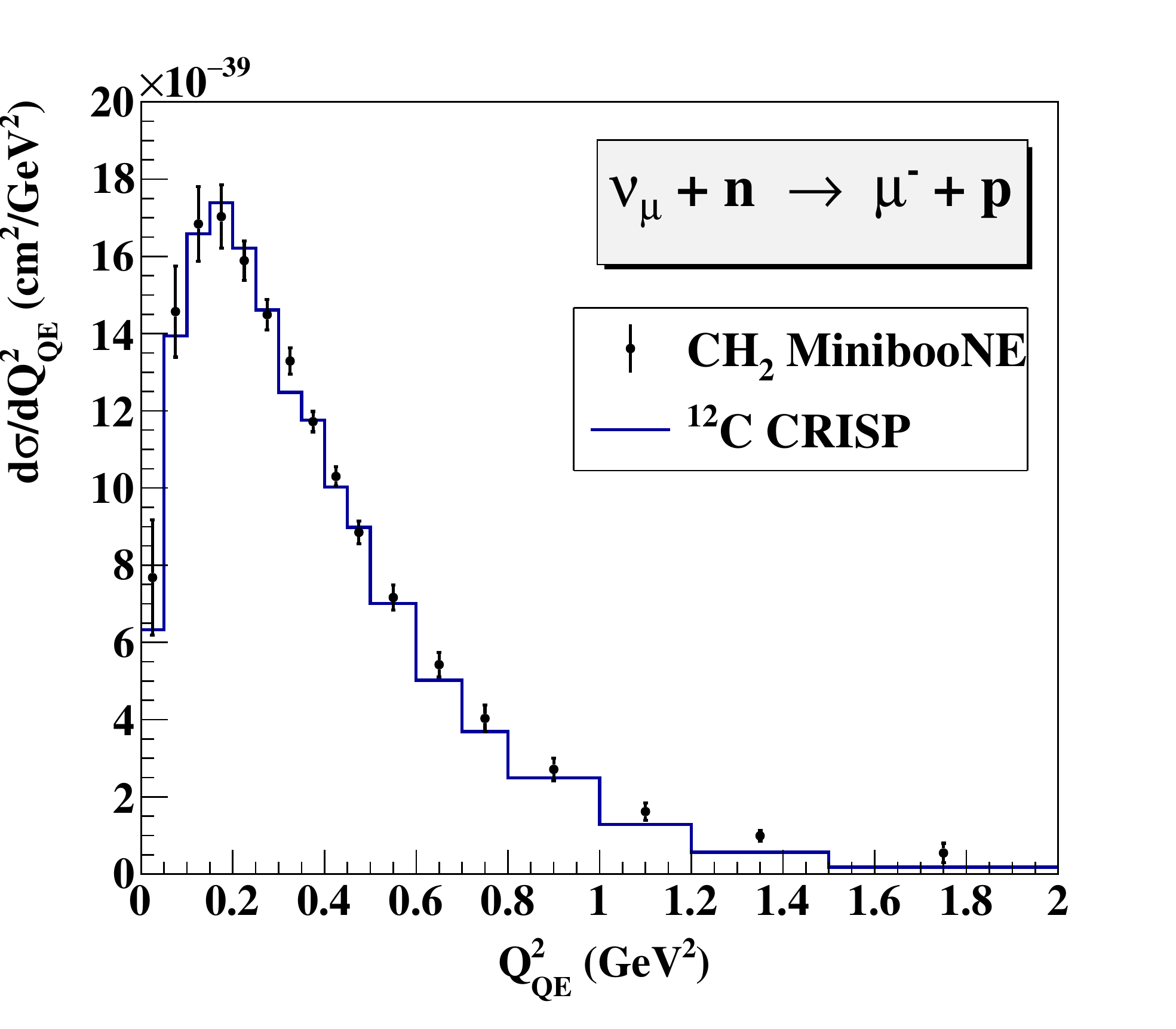}
	\raisebox{0.2\height}{\includegraphics[scale=0.3]{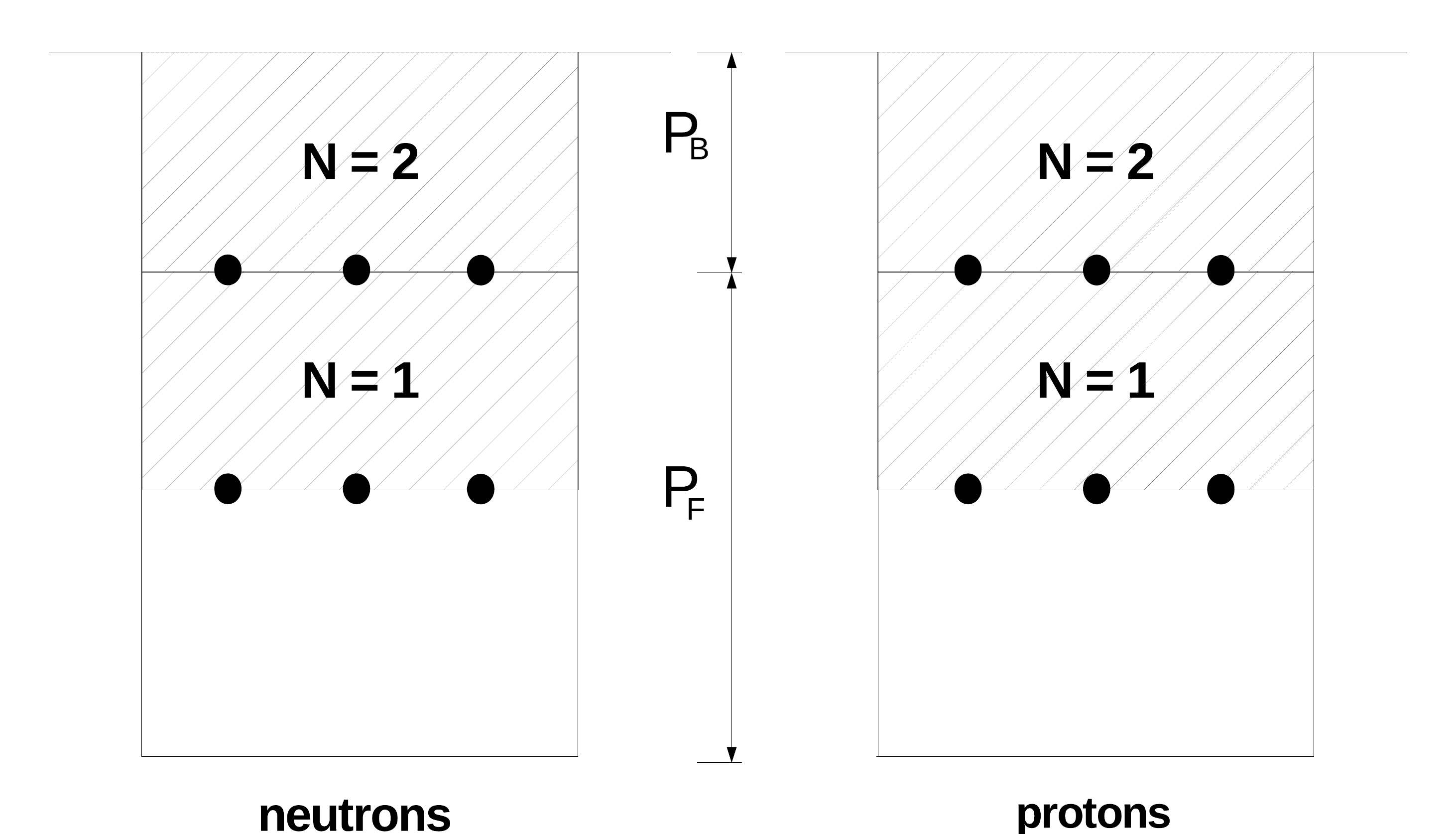}}
	\caption{Top: $d\sigma/dQ^2$ cross section per neutron for the channel $\nu_\mu + n \rightarrow \mu^- + p$ (CCQE) in reactions $\nu_\mu + ^{12}C$ (blue line) and $\nu_\mu + n$ (red line). The experimental data correspond to the reaction $\nu_\mu + CH_2$ \cite{aguilar-arevalo_first_2010}. Bottom: Shell 
	structure for $^{12}C$ as used in CRISP. $^{12}C$ is formed by two energy levels for neutrons 
	(protons). At the fundamental state, levels $N=1, \ 2$ are completely occupied by three neutrons 
	and three protons, respectively. $P_F$ is the Fermi momentum and $P_B$ is the momentum a nucleon 
	needs to be emitted.}
	\label{fig:4}
\end{figure}

The occupation number of each energy level (or shell) $n$ is defined by the number of combinations of quantum numbers which fulfill the condition $n_x^2+n_y^2+n_z^2=n^2$, times 2 due to spin. In an alternative model, each shell may be divided in two, one for nucleons with positive spin component and another for negative spin component. More realistic nuclear models take this into consideration after inclusion of spin-orbit coupling. In Fig. \ref{fig:4} (top), the total CCQE
cross section is presented using this model. One can observe that there is a good reproduction of the data. This is very likely due to a more effective Pauli blocking.

The model in Fig. \ref{fig:4} also allows to obtain a good reproduction of the double differential cross section of muon emission in the reaction $\nu_\mu + ^{12}C$ (Fig. \ref{fig:444}).

\begin{figure}[hbt!]
	\centering
	\includegraphics[scale=0.4]{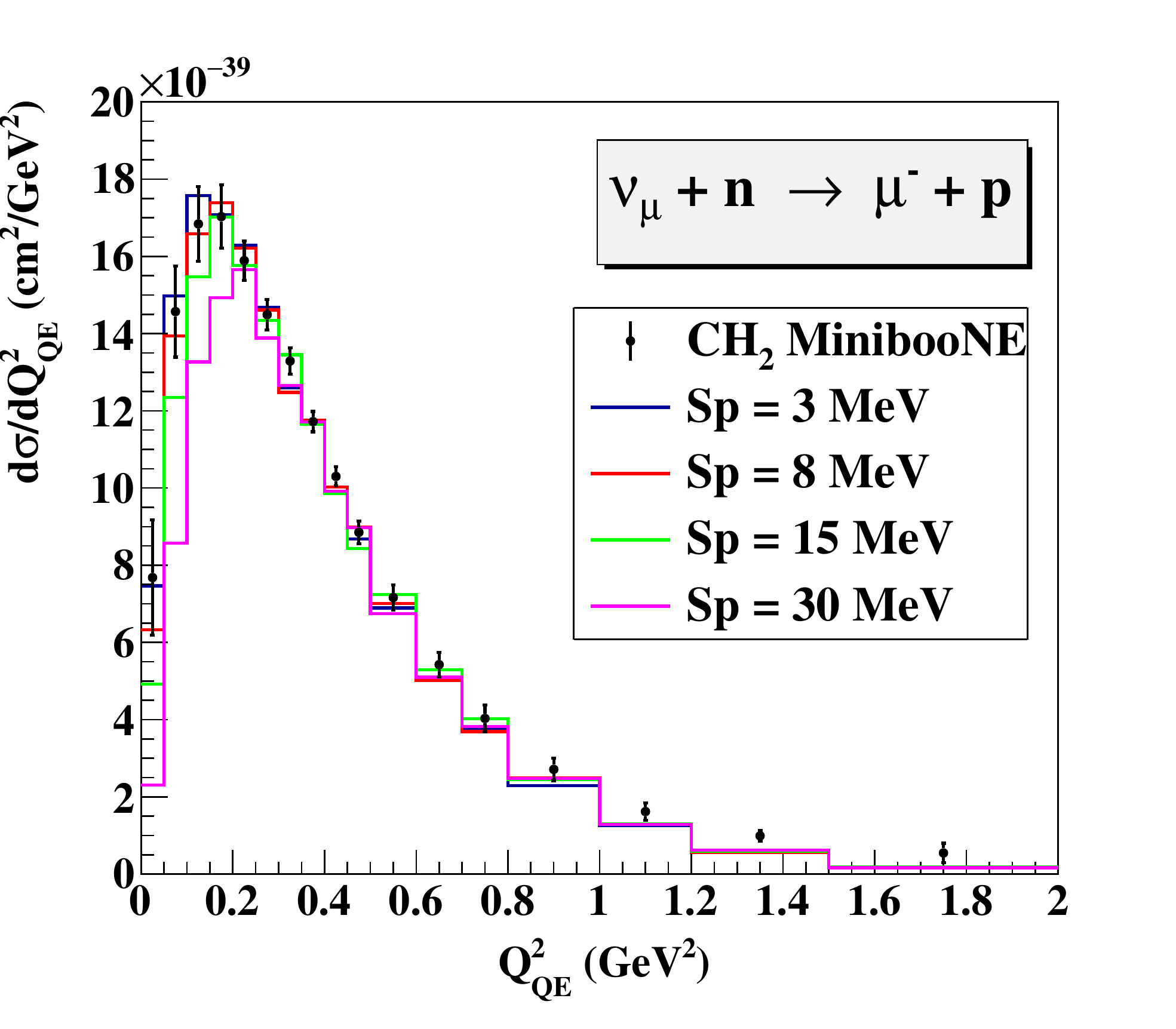}
	\caption{$d\sigma/dQ^2$ cross section per neutron for the reaction $\nu_\mu + n \rightarrow \mu^- + p$. The 
	experimental data correspond to the reaction $\nu_\mu + CH_2$ and were taken from 
	\cite{aguilar-arevalo_first_2010}. The simulations were performed for the reaction $\nu_\mu + ^{12}C$.}
	\label{fig:3aa}
\end{figure}

In the figure \ref{fig:3aa} there has been represented the cross section $d\sigma/d Q^2$ depending on the proton separation energy of the target nucleus. Since the CRISP model was initially developed to study nuclear reactions in heavy nucleus, it uses as nucleon separation energy the binding energy per nucleon S = B/A = 8 MeV. In the case of light nuclei, other values of separation energy offers a better match between our simulations and the experimental data. For example, in the reaction $\nu_\mu + CH_2$, the best result of $d\sigma/dQ^2$ (CCQE channel) is obtained when $S_p = 3 \ MeV$. When the neutrino interacts with $^{12}C$ through the CCQE channel, the target neutron is transformed in a proton and therefore the $^{12}C$ is transformed in $^{12}N$. The proton separation energy of $^{12}N$ is $0.6 \ MeV$ \cite{wang_ame2016_2017}, which explains why our best results are obtained for $S_p = 3 \ MeV$.
Figure \ref{fig:444} shows muon neutrino double-differential cross section for CCQE scattering on hydrocarbon in terms of kinetic energy and scattering angle of the emitted muon. Can be observed a good agreement of CRISP results with the experimental data. It is important to note that we have made the calculations in $^{12}C$, as the two free protons of $CH_2$ do not interact with the incident muon neutrino through CCQE channel.
\begin{figure}[hbt!]
	\centering
	\includegraphics[scale=0.42]{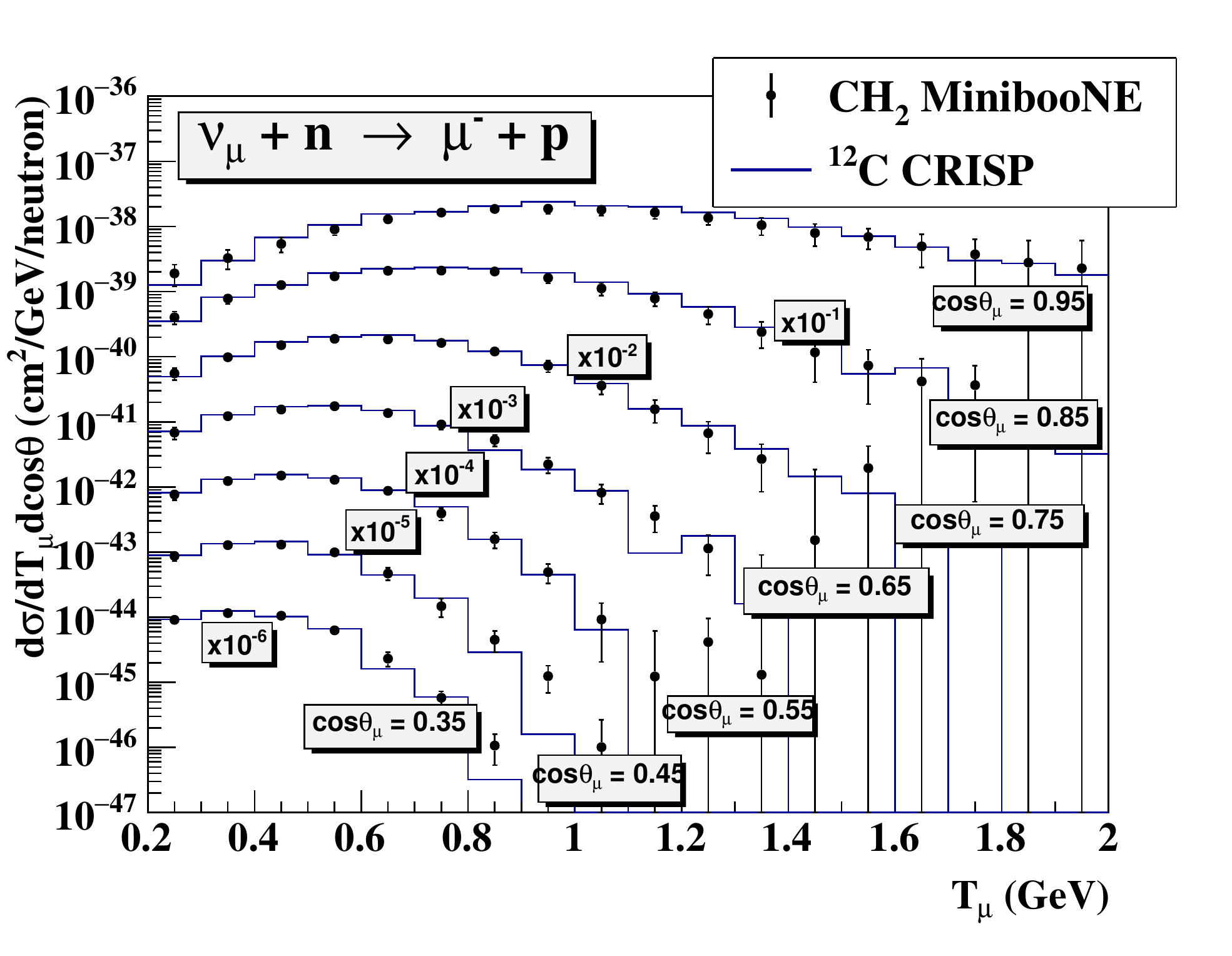}
	\includegraphics[scale=0.42]{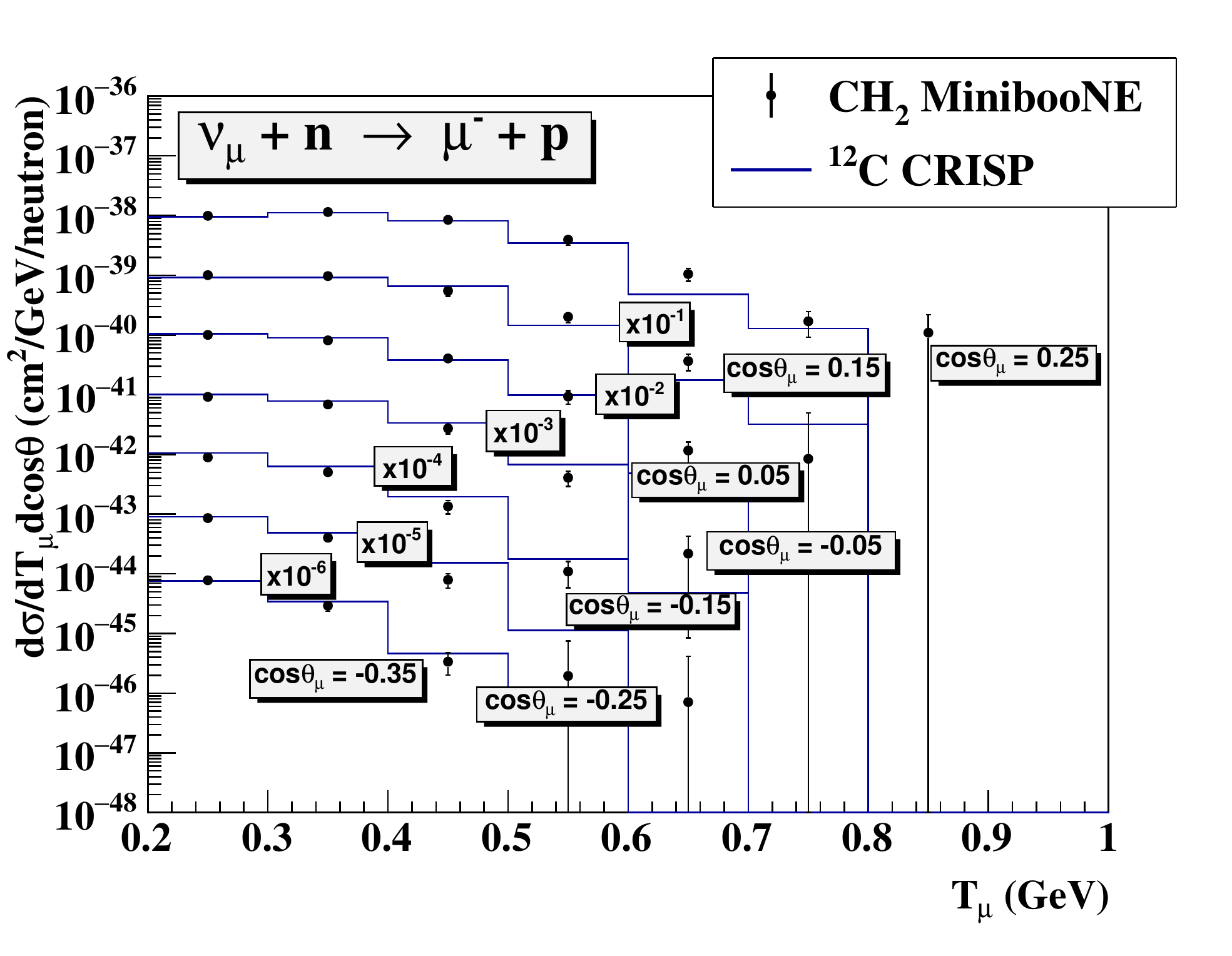}
	\caption{Double differential cross section $d\sigma /dT_\mu d\cos \theta_\mu$ of emission of $\mu^-$ 
	of the CCQE channel. The reaction $\nu_\mu + ^{12}C$ was calculated using the shell structure of 
	Fig. \ref{fig:4}. The experimental data were taken from \cite{aguilar-arevalo_first_2013}.}
	\label{fig:444}	
\end{figure}
\begin{figure}[hbt!]
	\centering
	\includegraphics[scale=0.4]{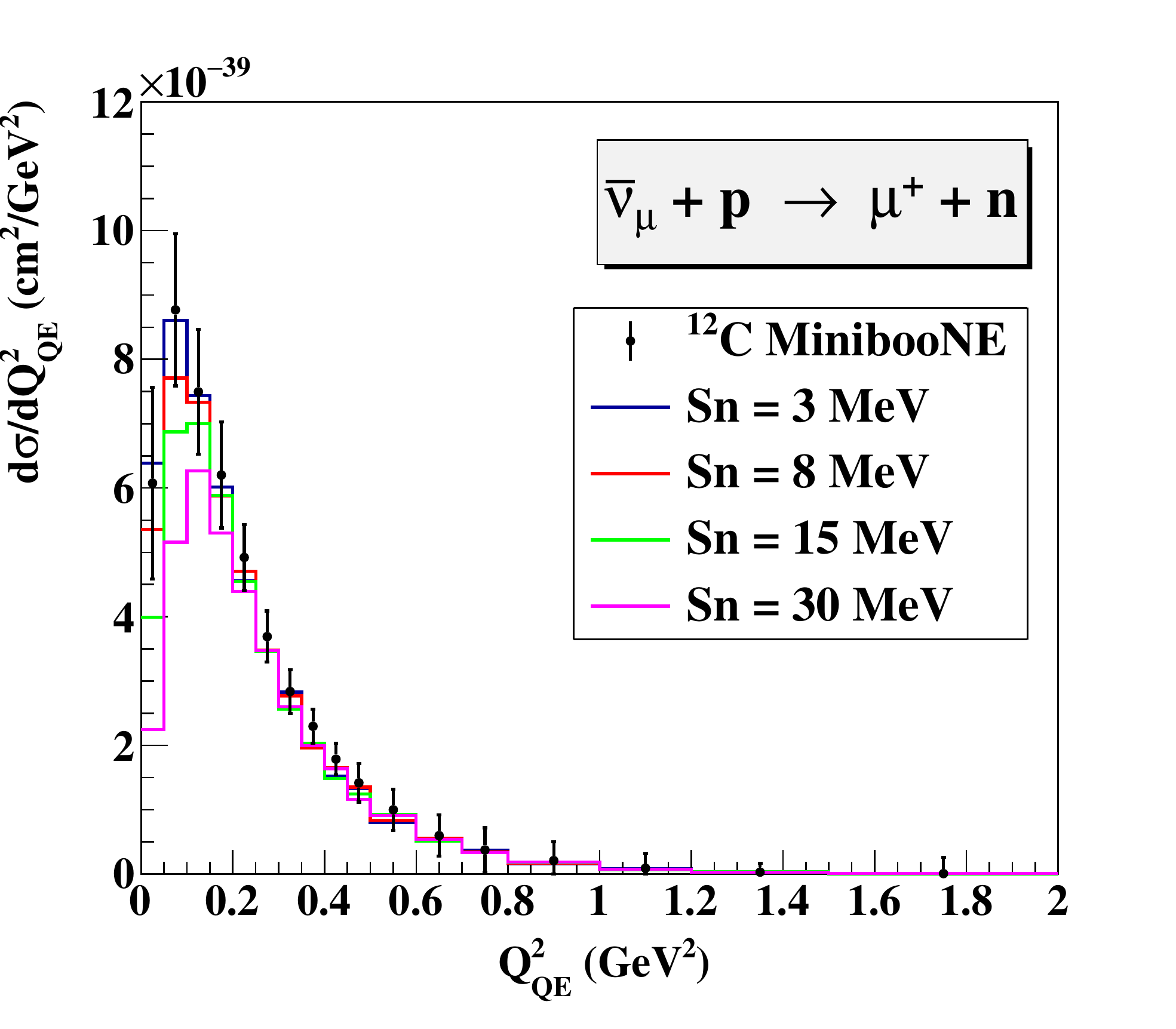}
	\includegraphics[scale=0.4]{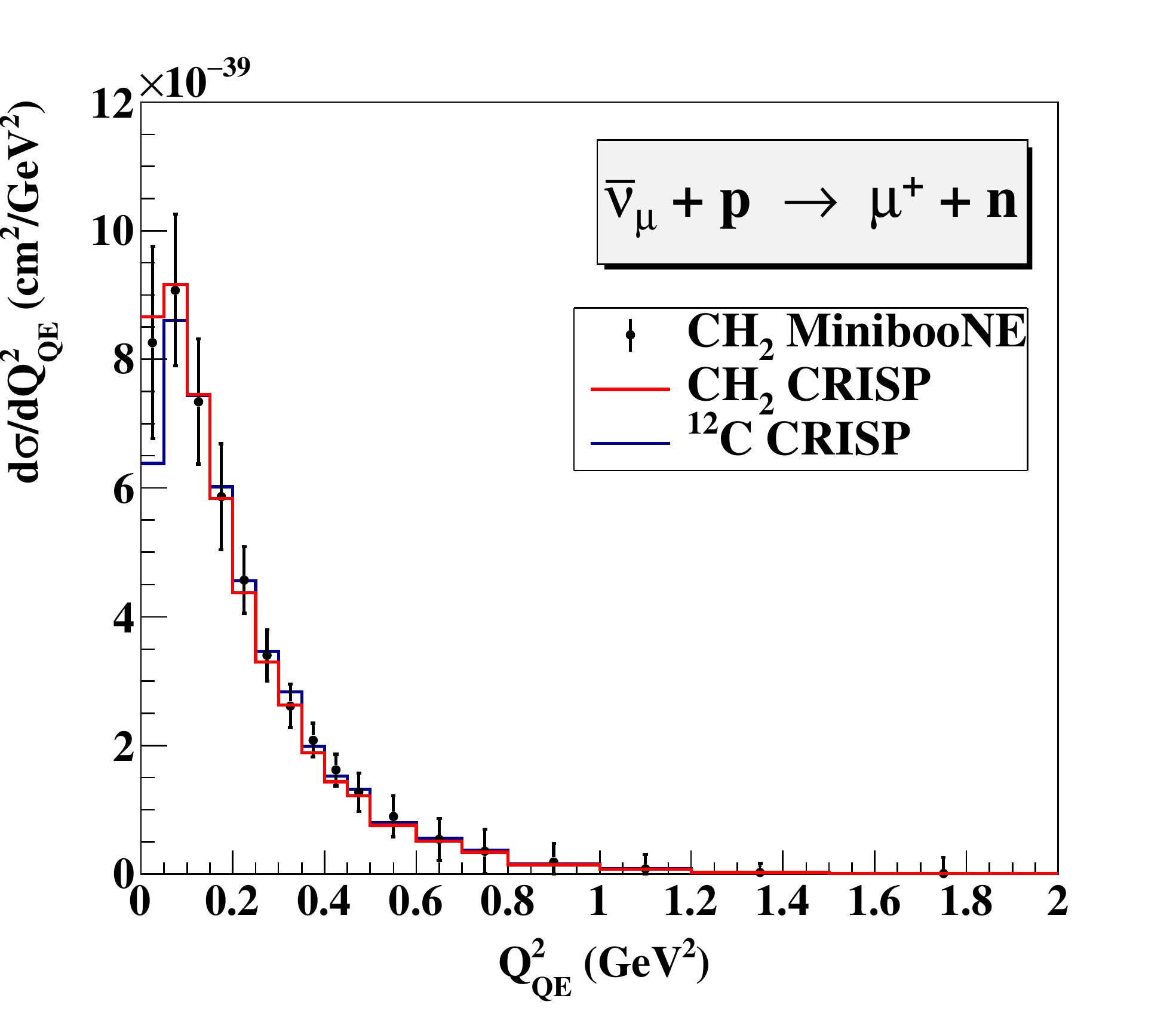}
	\caption{NCE $d\sigma/dQ^2$ cross section per proton for the reactions $\bar{\nu}_\mu + ^{12}C$ (top) and $\bar{\nu}_\mu + CH_2$ (bottom).  $S_n$ is the neutron separation energy used to build the target nucleus. Experimental data extracted from \cite{aguilar-arevalo_first_2010}.} 
	\label{fig:333}
\end{figure}
Figure \ref{fig:333} (top) presents the $d\sigma/dQ^2$ cross section of the $\bar{\nu}_\mu + p \rightarrow \mu^+ + n$ reaction, for different values of nucleon separation energy. In this case, when the antineutrino interacts with a proton, the $^{12}C$ is transformed in $^{12}B$, which has a proton separation energy equal to $3.4 \ MeV$. This explains why the best results of CRISP simulations are obtained with $S_N = 3 \ MeV$. The red line of figure \ref{fig:333} (bottom) represents the $d\sigma/dQ^2$ cross section of $\bar{\nu}_\mu + CH_2$ reaction.
In figure \ref{fig:333} (bottom), we have the $\frac{d\sigma}{dQ^2}$ cross section for the $\bar{\nu}_\mu + CH_2 \rightarrow \mu^+ + n$ reaction (red line) and the $\nu_\mu + ^{12}C$ reaction (blue line). The red line is calculated as the anti-neutrino cross section on $^{12}C$ plus the anti-neutrino cross section on two free protons. The main difference between the two reactions is in the $0 < Q^2 < 0.2$ $GeV^2$ interval, where the $CH_2$ cross section is higher than the $^{12}C$ cross section. That's because the Pauli exclusion principle is not present when the anti-neutrino interacts with any of the free protons on $CH_2$. In both studied reactions can be observed a good agreement with the experimental data.
\begin{figure}[hbt!]
	\centering
	\includegraphics[scale=0.4]{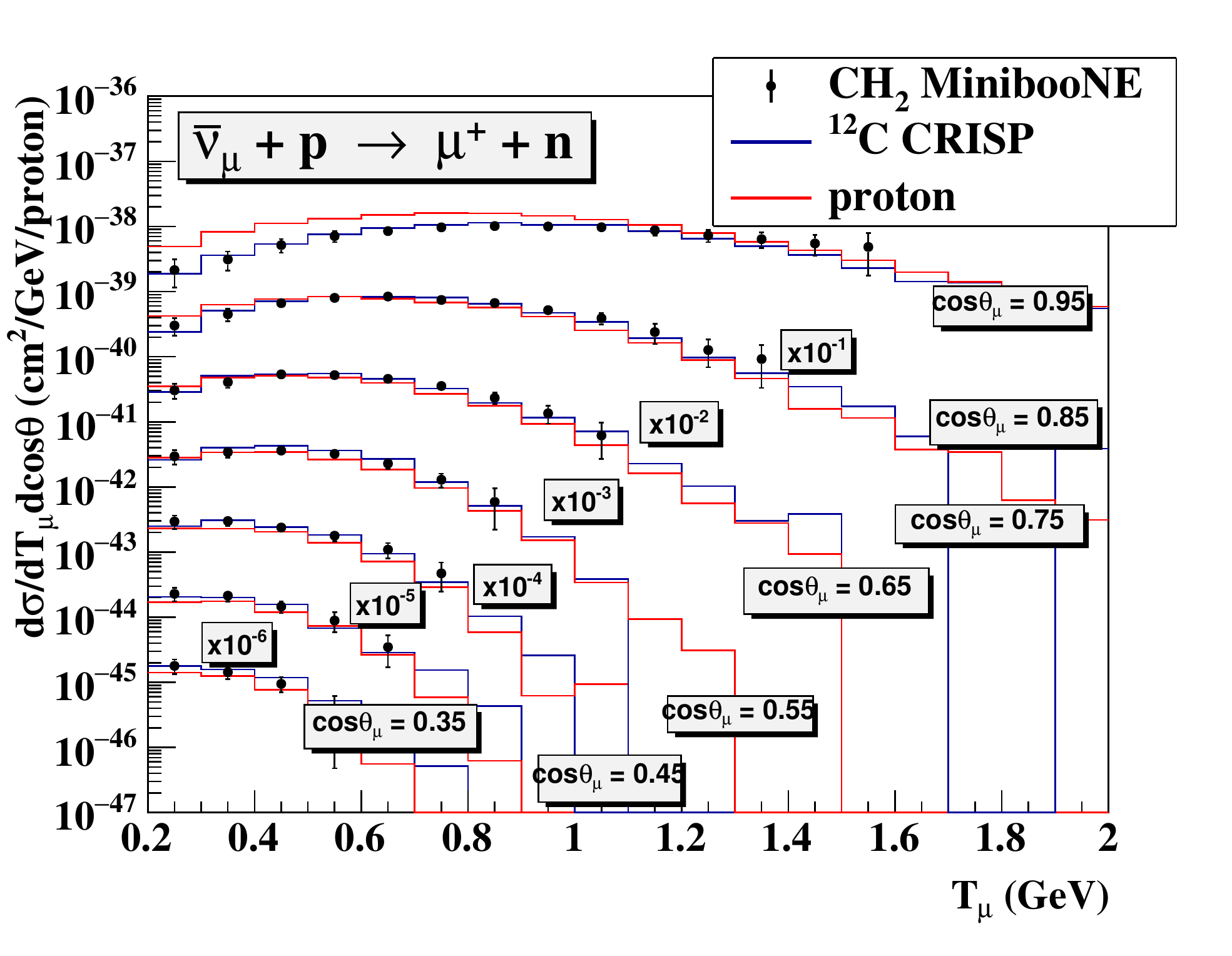}
	\includegraphics[scale=0.4]{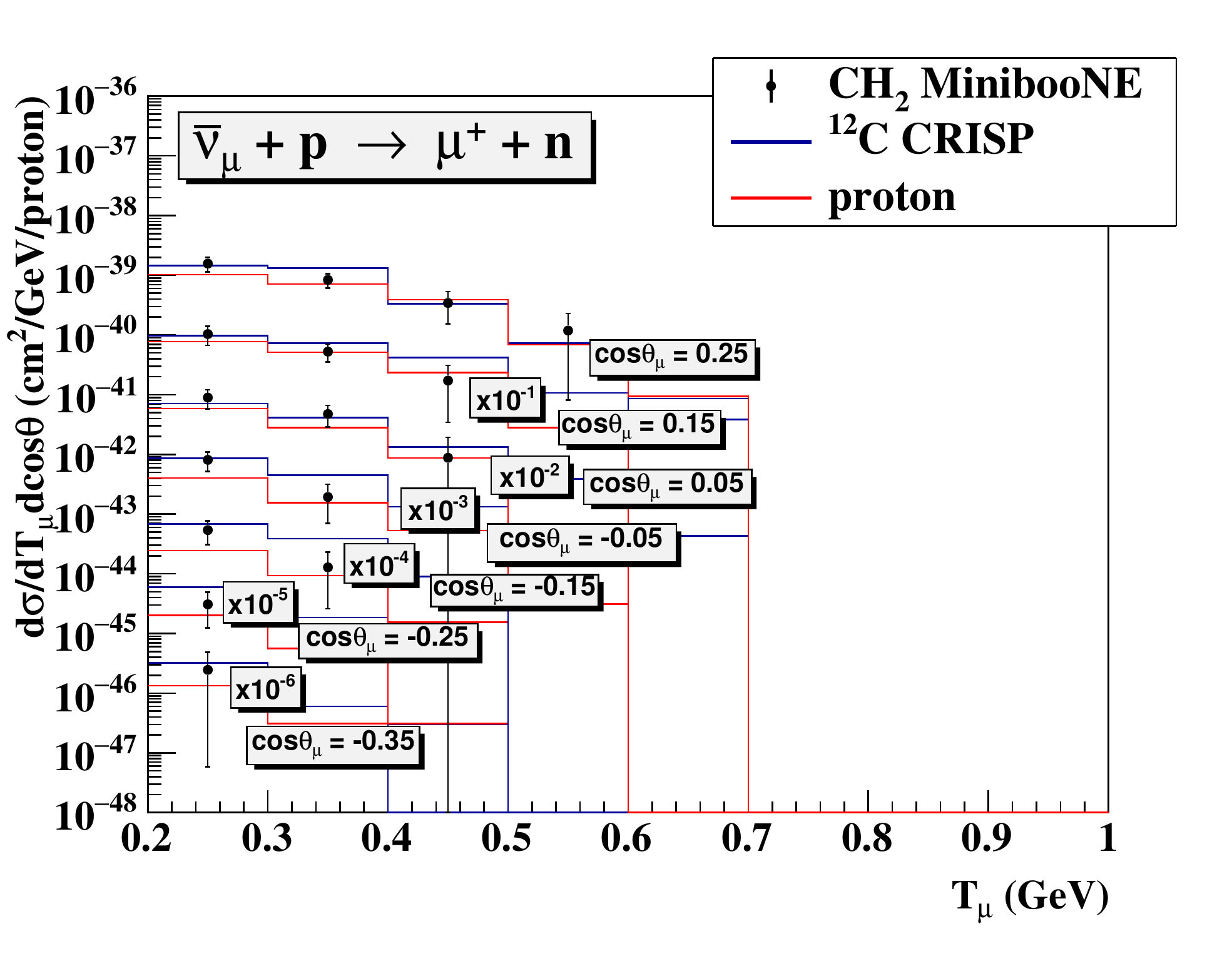}
	\caption{Double differential cross section $d\sigma /dT_\mu d\cos \theta_\mu$ of emission of $\mu^+$ 
	of the CCQE channel for the reaction $\bar{\nu}_\mu + ^{12}C$. Experimental data extracted from \cite{aguilar-arevalo_first_2010}.}
	\label{fig:334}
\end{figure}
A comparison between the muon anti-neutrino double-differential cross section on carbon and free protons is shown in figure \ref{fig:334}. The cross section on free protons is higher than on $^{12}C$ for the smallest scattering angles. That is because the Pauli exclusion principle limits the interactions with less energy transfer to the target nucleon, and therefore the reactions where the muon spreads forward. In this scattering, the anti-neutrinos with lower energy transfer less energy to the target nucleon, and there is a higher probability that the Pauli exclusion principle blocks the reaction. For the larger muon scattering angles, it is the opposite; the cross section on $^{12}C$ is higher than on free protons, in this case, higher energy is transferred to the target nucleon, and therefore, Pauli's blocking is less restrictive. Now, the difference between the two reactions is due to the fermionic movement of the nucleons on $^{12}C$.

\subsection{Quasi-elastic neutral current channel. Axial and strange form factors.}

For the measurement of the neutral current elastic channel (NCE), are selected the events with a muon and no mesons emitted in the intranuclear cascade (NCE like). According to  \cite{the_miniboone_collaboration_measurement_2010}, the magnitude $Q^2$ is determined experimentally from the measurement of the total kinetic energy of the emitted nucleons assuming the target nucleon at rest:
\begin{equation}
    Q_{E}^2 = 2m_N T = 2 m_N \sum_{i}T_i,
    \label{Q_E}
\end{equation}
where $T$ is the sum of the kinetic energy $T_i$ of each emitted nucleon.



As well as the CCQE channel, the following NCE events will be lost and measured as NCRes (neutral current resonance formation):
\begin{equation}
    \begin{tikzcd}[row sep = tiny, column sep = tiny]
        \nu_\mu (\bar{\nu}_\mu) + N \rar & \nu_\mu (\bar{\nu}_\mu) + N \arrow[d, start anchor={[xshift=4ex, yshift=1ex]},
end anchor={[xshift=-0.8ex,yshift=0ex]}] \\
        & N + N \rar & N + N^* \arrow[d,start anchor={[xshift=2ex, yshift=1ex]},
end anchor={[xshift=-0ex,yshift=0ex]}] \\
        & &  N^* \rar & N + \pi
    \end{tikzcd} 
    \label{eq.NCQE3}
\end{equation}
and the following NCRes channel events will be reported as NCE:
\begin{equation}
    \begin{tikzcd}[row sep = tiny, column sep = tiny]
        \nu_\mu (\bar{\nu}_\mu) + N \rar & \nu_\mu (\bar{\nu}_\mu) + N^* \arrow[d, start anchor={[xshift=4ex, yshift=1ex]},
end anchor={[xshift=1.0ex,yshift=0ex]}] \\
        & N^* \rar & N + \pi \arrow[d, start anchor={[xshift=1ex, yshift=0ex]},
end anchor={[xshift=-0.8ex,yshift=0ex]}] \\
        & & \pi + N \rar & N + N.
    \end{tikzcd} 
    \label{eq.NCQE4}
\end{equation}

This resonant contribution (equation \ref{eq.NCQE4}) is considered in the MiniBooNE experiment and taken from the "NCE like" cross section to obtain the NCE cross section. In figure 62, we have the NCE cross section calculated by CRISP for the $\nu_\mu + ^{12}C$ (top) and $\bar{\nu}_\mu + ^{12}C$ (bottom) reactions.

\begin{figure}[hbt!]
	\centering
	\includegraphics[scale=0.4]{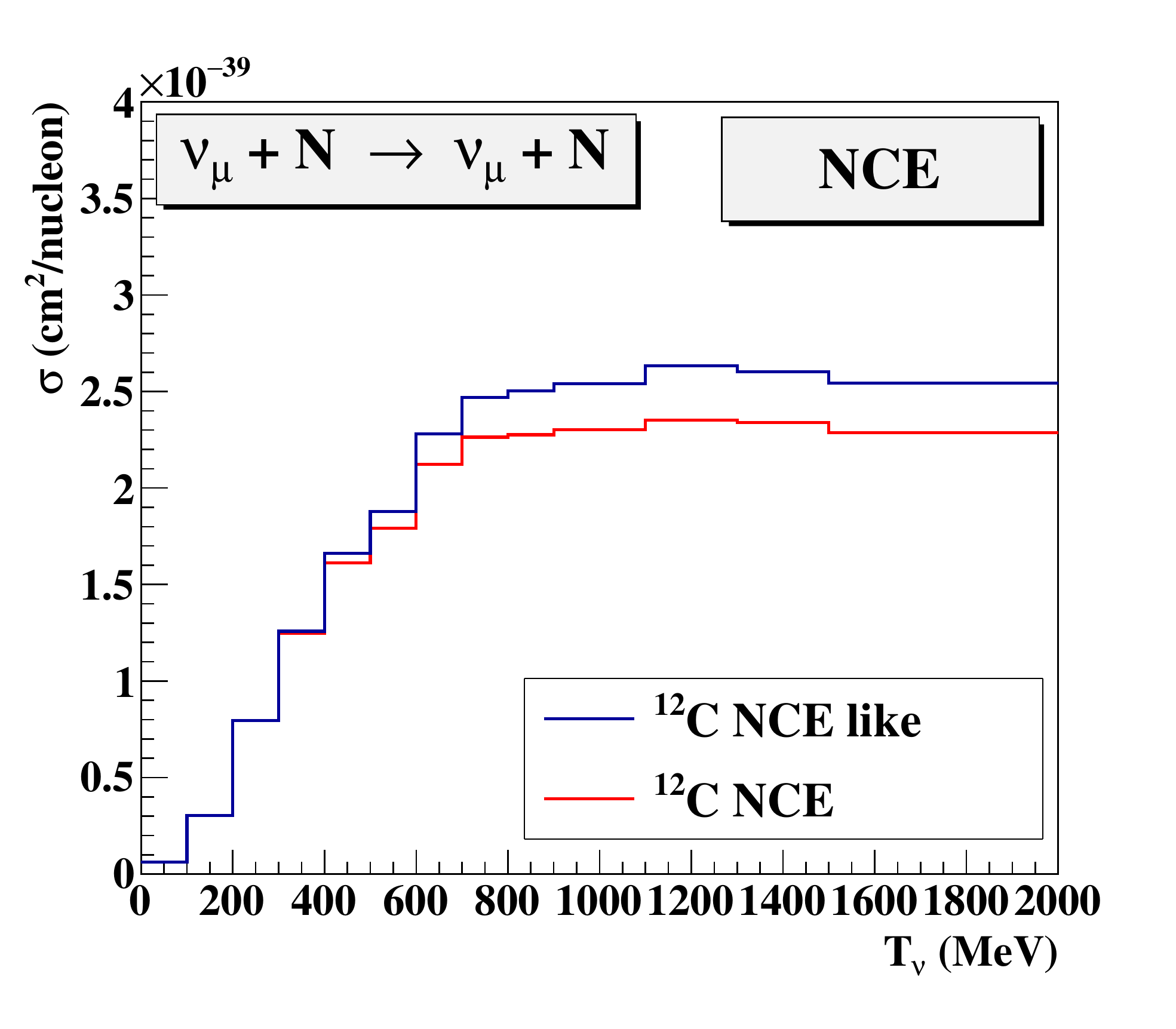}
	\includegraphics[scale=0.4]{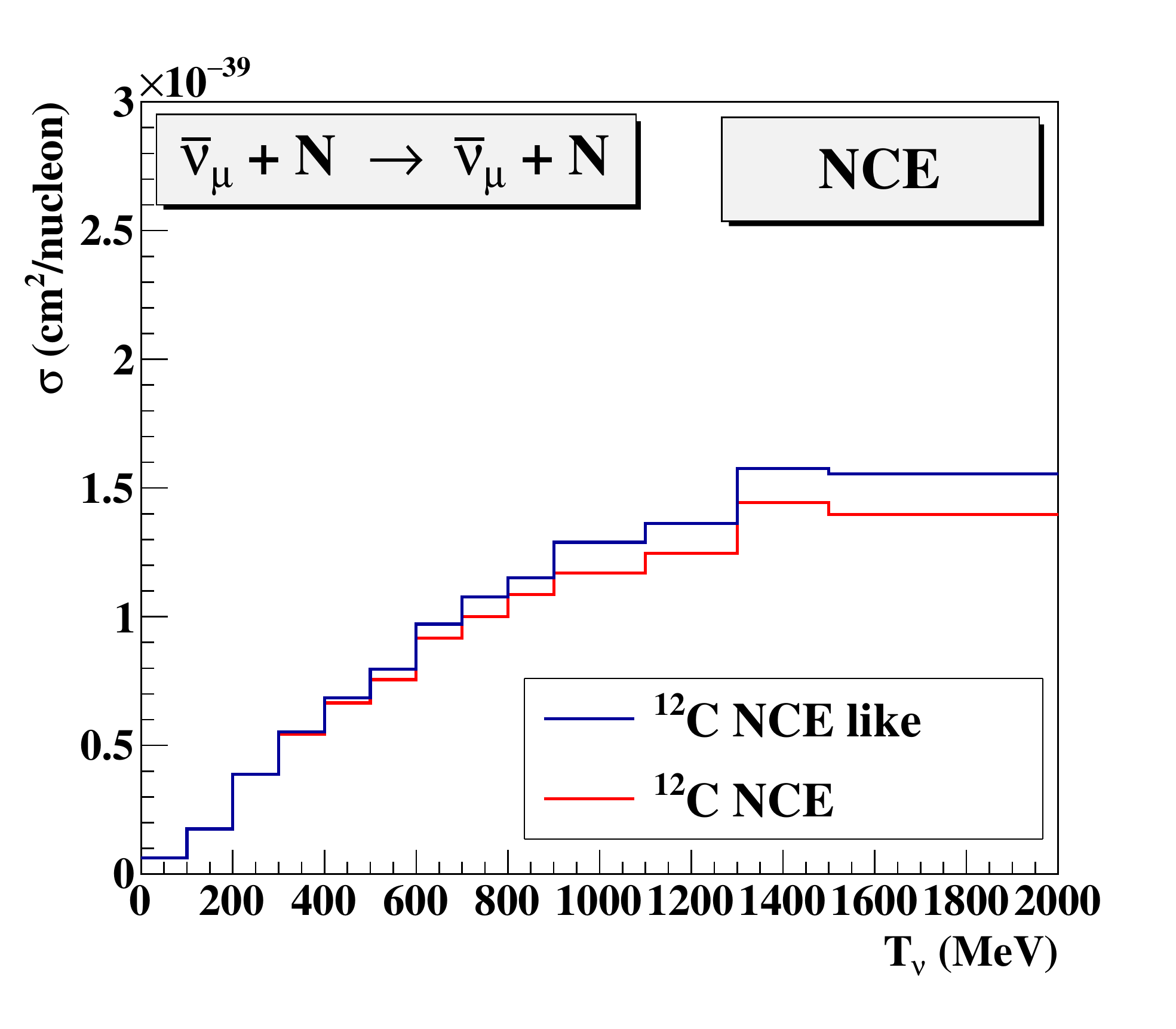}
	\caption{cross section per nucleon for the NCE channel for the reactions $\nu_\mu + ^{12}C$ (top) and $\bar{\nu}_\mu + ^{12}C$ (bottom).}
	\label{fig:62}
\end{figure}

\begin{figure}[hbt!]
	\centering
	\includegraphics[scale=0.4]{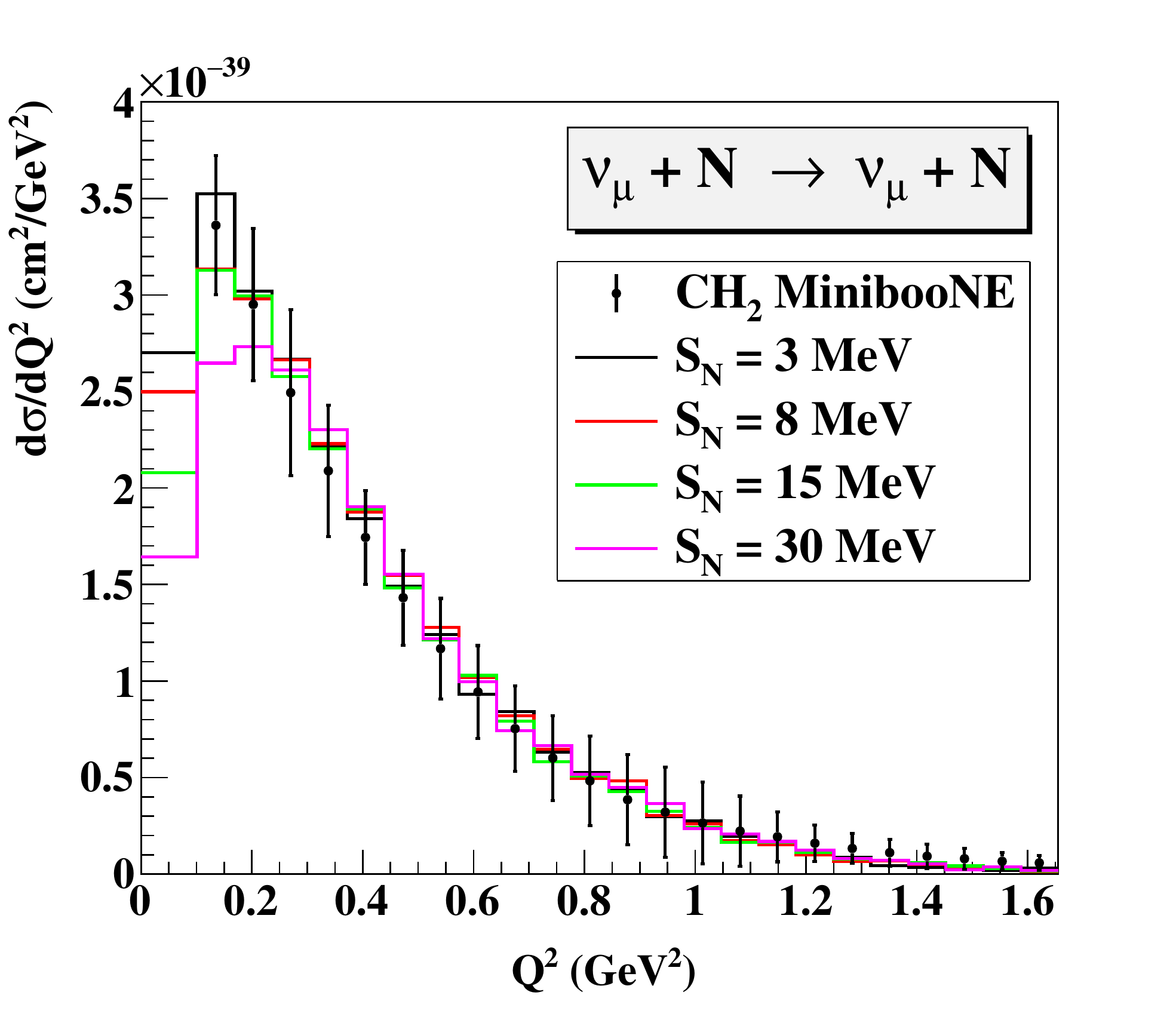}
	\includegraphics[scale=0.4]{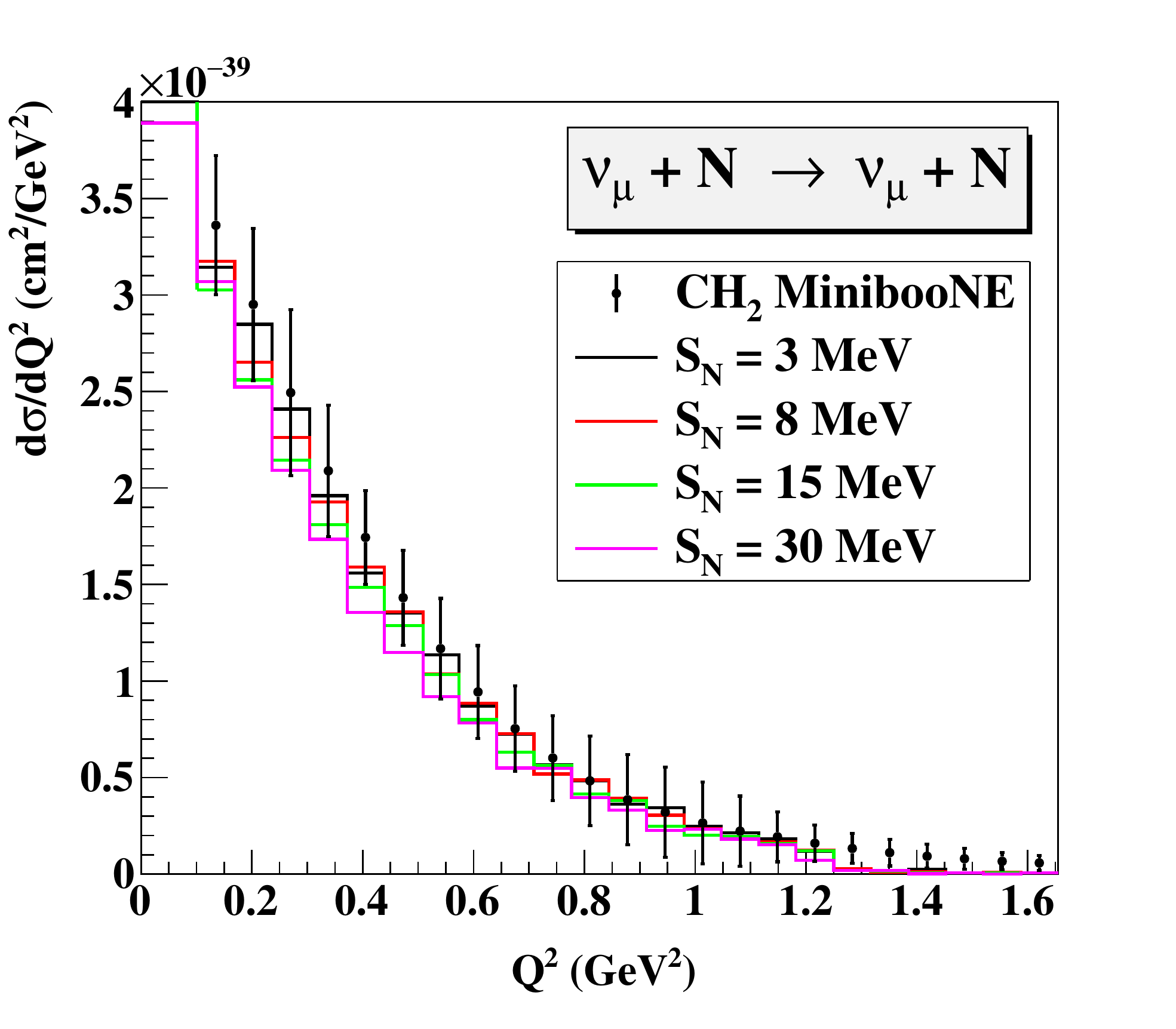}
	\caption{$d\sigma/dQ^2$ cross section per nucleon for the NCE channel for the reaction $\nu_\mu + CH_2$. Top: $Q^2$ computed as $Q^2 = -(p_\nu^\prime - p_\nu)^2$ (see text). Bottom: $Q^2$ computed as equation \ref{Q_E}.  $S_N$ is the nucleon separation energy used to build the target nucleus. Experimental data were taken from \cite{the_miniboone_collaboration_measurement_2010-1}}
	\label{fig:62a}
\end{figure}

Since CRISP code allows knowing the four-momentum of each involved particle in the reaction, it is possible to calculate the exact value of $Q^2$ using the formula $Q^2 = -(p_\nu^\prime - p_\nu)^2$, where $p_\nu$ and $p_\nu^\prime$ are the incident and scattered neutrino fourth-momentum respectively. The  $\nu_\mu + CH_2$ ($\bar{\nu}_\mu + CH_2$) differential cross section for that calculation, and different $S_N$, is shown in figure \ref{fig:62} top (figure \ref{fig:62a} top). It can be observed that the value that best reproduces the experimental data is $S_N = 15$ MeV, which is in agreement with the separation energy values for $^{12}C$, where $S_n = 18.7$ MeV and $S_p = 16.0$ MeV \cite{wang_ame2016_2017}.

\begin{figure}[hbt!]
	\centering
	\includegraphics[scale=0.4]{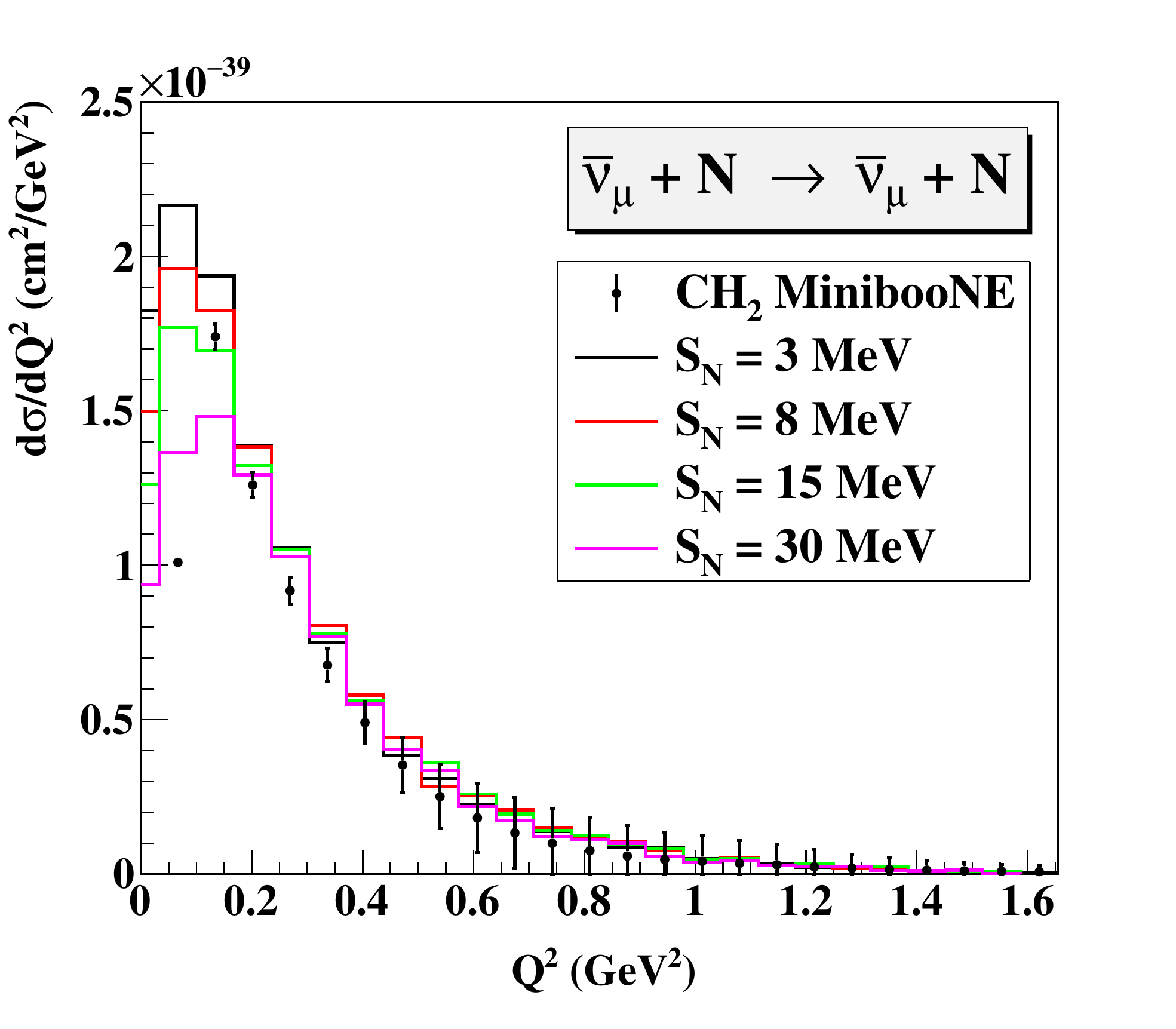}
	\includegraphics[scale=0.4]{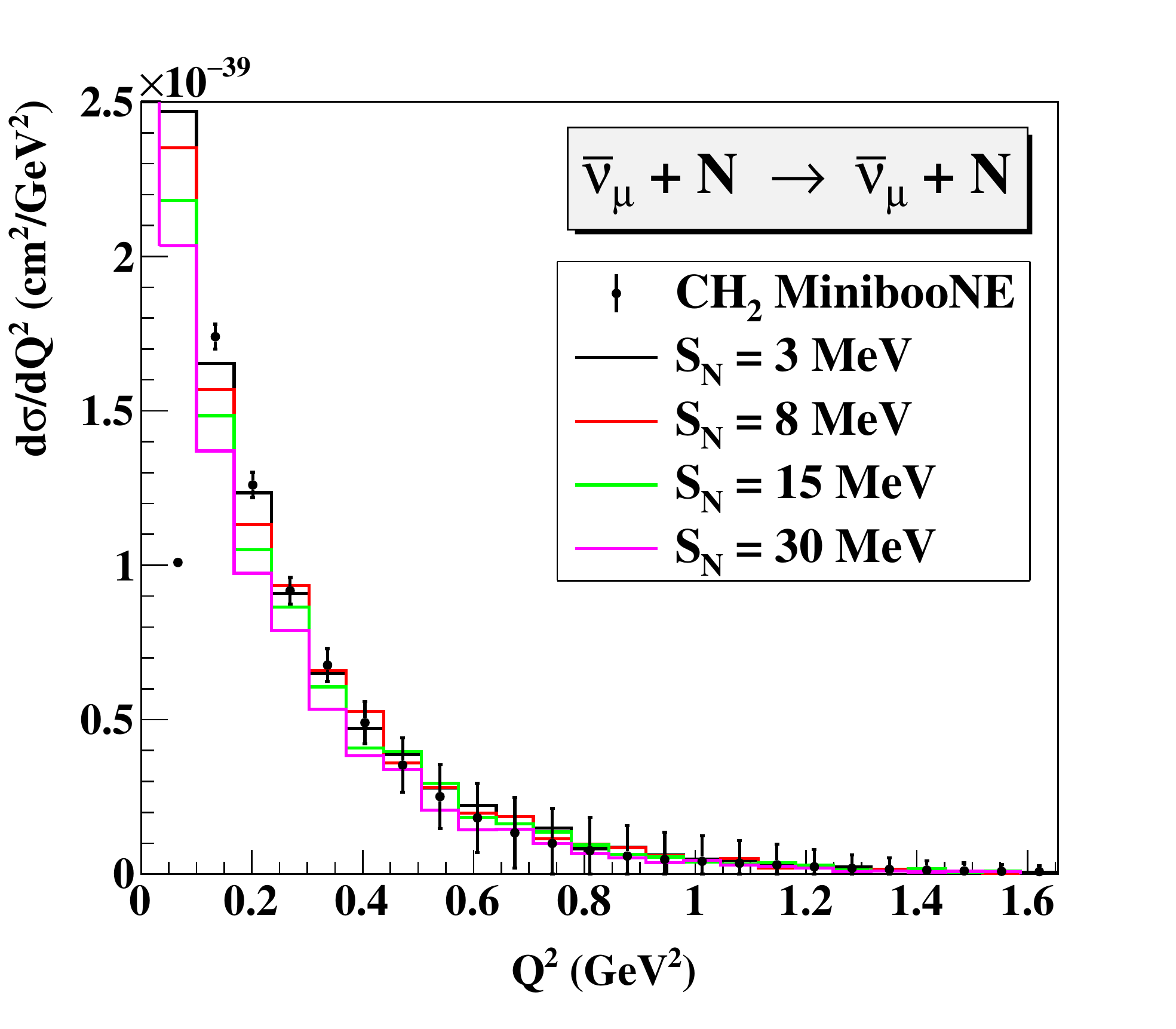}
	\caption{$d\sigma/dQ^2$ cross section per nucleon for the NCE channel for the reaction $\bar{\nu}_\mu + CH_2$. Top: $Q^2$ computed as $Q^2 = -(p_\nu^\prime - p_\nu)^2$ (see text). Bottom: $Q^2$ computed as equation \ref{Q_E}.  $S_N$ is the nucleon separation energy used to build the target nucleus. Experimental data were taken from \cite{aguilar-arevalo_measurement_2015}}
	\label{fig:62b}
\end{figure}

When calculating $d\sigma/dQ^2$ using relation \ref{Q_E}, we have a good agreement between calculation and experiment (see bottom figures \ref{fig:62a} and \ref{fig:62b}). In range $Q^2 < 0.4 \ GeV^2$, CRISP underestimates the experimental data to the point that it cannot reproduce the peak shape cross section. To explain that, it is necessary to understand how the NUANCE code works \cite{casper_nuance_2002}. NUANCE is the Monte Carlo simulation model used to process the experimental measurements. It was used to assess the efficiency of the detectors and to obtain the background processes for the "NCE-like" cross section (equation \ref{eq.NCQE4}). Thus, the results obtained in the experiment depend on the NUANCE model.

\begin{figure}[hbt!]
	\centering
	\includegraphics[scale=0.35]{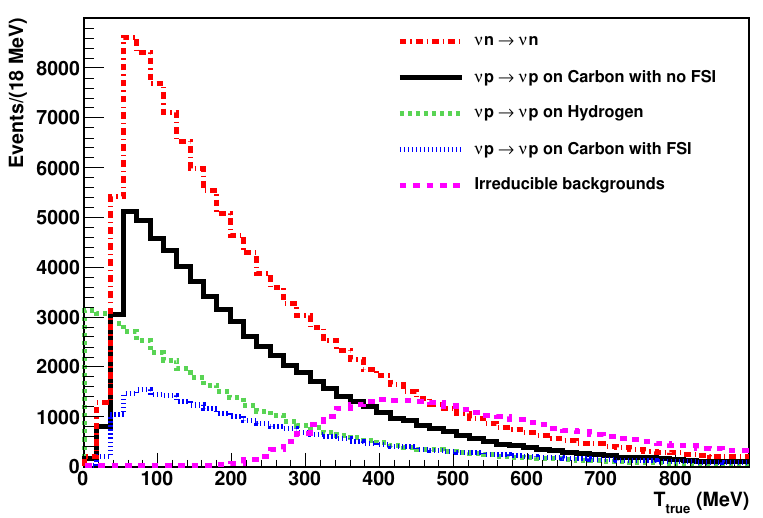}
	\caption{Kinetic energy histograms of the emitted nucleons that contribute to the "NCE like" cross section, calculated with NUANCE code for the reaction $\nu_\mu + ^{12}C$. Figure extracted from \cite{perevalov_neutrino-nucleus_2009}.}
	\label{fig:62c}
\end{figure}

Next, we will analyze some relevant aspects of the NUANCE model; for this, we start from the NUANCE calculation of kinetic energy distribution of the nucleons emitted in different processes contributing to the NCE channel (figure  \ref{fig:62c}). The green line in figure  \ref{fig:62c} represents the NCE cross section of reaction on hydrogen. This curve does not have a peak shape because the Pauli principle is not present on free isolated protons. An important fact in these results is that the exclusive carbon neutrino cross section has a peak at $T\approx90$ MeV (black line), even considering that the nucleons detected in the experiment were already emitted during the intranuclear cascade. When a reaction is not allowed by the Pauli exclusion principle, it is because the resulting nucleons are energetically confined to the nuclear potential and therefore do not have enough energy to escape from the nucleus. In other words, the nucleons emitted in the intranuclear cascade should not be influenced by the Pauli block, as it is obtained in the CRISP calculations.

In the NUANCE model, to take into account the nuclear effect, the following correction is applied to the cross section:

\begin{equation}
    \sigma = (1 - D/N)\sigma_{free},
\end{equation}
where $N$ is the number of nucleons, $\sigma_{free}$ is the  neutrino -  nucleon free cross section  and D is:
\begin{equation}
     D = \left\{
	       \begin{array}{ll}
		 \frac{A}{2}\left( 1 - \frac{3}{4} \frac{|\vec{q}|}{p_F} + \frac{1}{16}(\frac{|\vec{q}|}{p_F})^3\right)      & \mathrm{if\ } |\vec{q}| < 2p_F \\
		 0 & \mathrm{if\ } |\vec{q}| > 2p_F 
	       \end{array}
	     \right.,
	     \label{correction}
\end{equation}
where, $A$ is the mass number, $p_F$ the Fermi momentum, and $q$ the transferred momentum to the target nucleus.

Correction  \ref{correction} influences the free cross section until where Q represents the emitted nucleons with $T = 90$ MeV, and for that reason, the exclusive carbon neutrino cross section has a peak at $T\approx90$ MeV (black line of figure  \ref{fig:62c}). In effect, let us consider a valence nucleon with $P = P_F = 220 \ MeV/c$ and an NCE interaction with the maximum transferred momentum so that the cross section is modified by equation \ref{correction}, that is, $q = 2P_F = 440 \ MeV/c$.  In that case, the final nucleon momentum can be $P = 440+220 = 660 \ MeV/c$. The kinetic energy for that momentum is $T \approx  208 \ MeV$ (inside the nucleus). Considering a nuclear potential of $V = 40 \ MeV$, the nucleon kinetic energy offside the nucleus is $T \approx 208 - 40 = 168 > 90 \ MeV$ (when it is emitted). 



\begin{figure}[hbt!]
	\centering
	\includegraphics[scale=0.4]{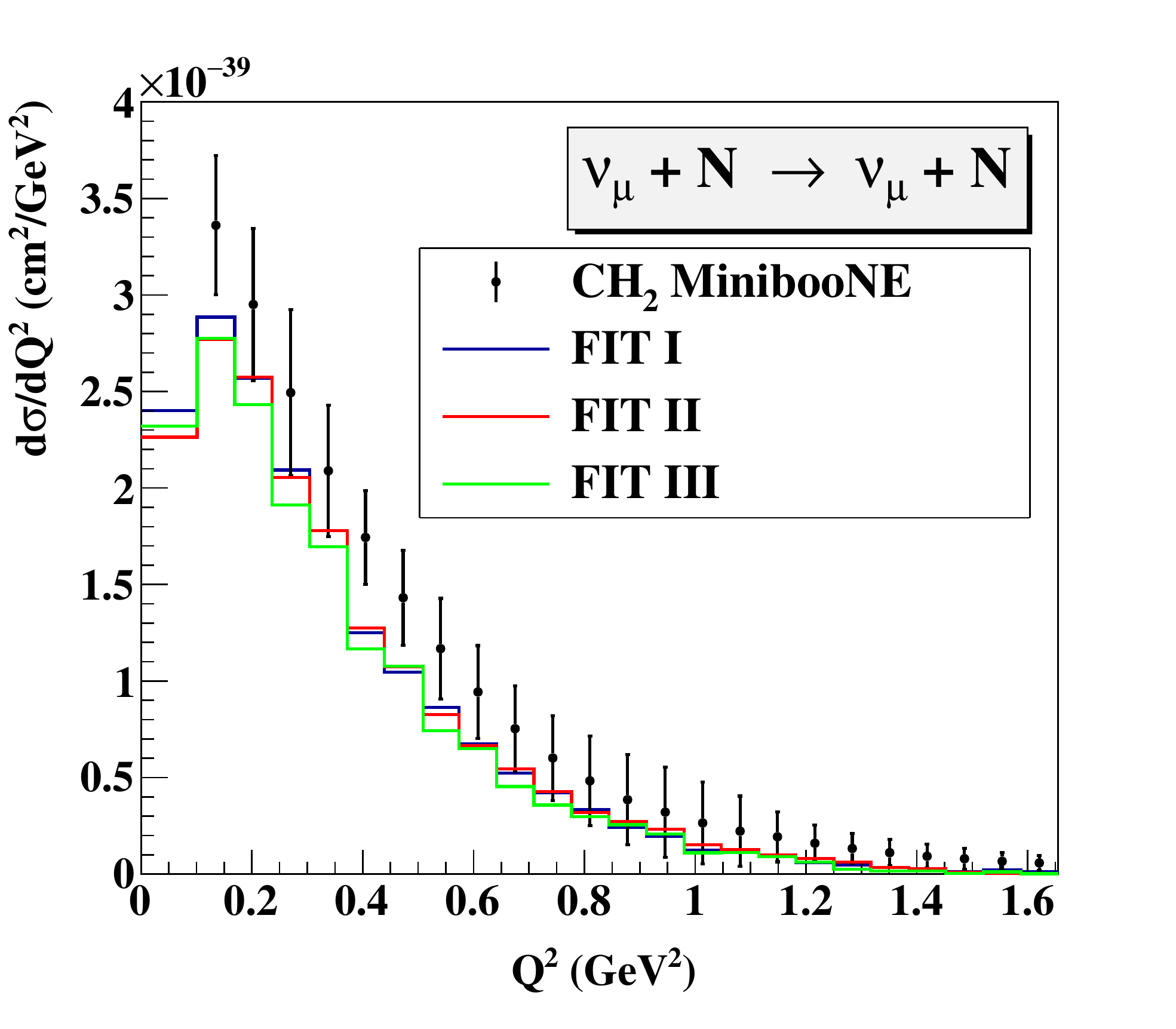}
	\includegraphics[scale=0.4]{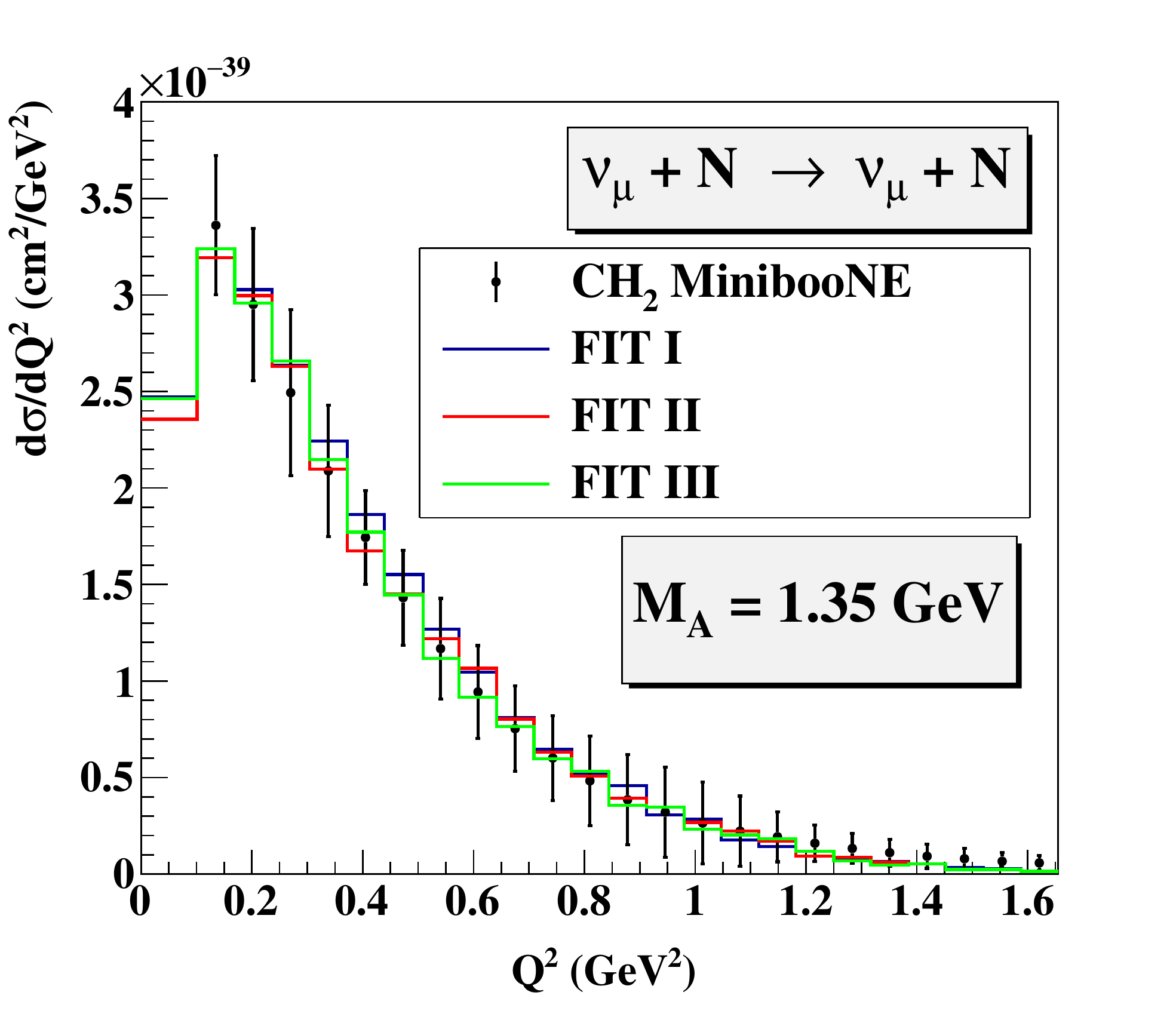}
	\caption{$d\sigma /dQ^2$ cross section for the NCE channel for the reaction $\nu_\mu + CH_2$, with parameterization from table \ref{tab:param}. Top: Parameter $M_A$ from table \ref{tab:param}; Bottom: Parameter $M_A = 1.35 \ GeV$. Experimental data extracted from \cite{the_miniboone_collaboration_measurement_2010-1}.}
	\label{fig:6}
\end{figure}

NCE interaction is interesting for the study of axial ($F_A$) and strange ($F_1^S$, $F_2^S$, $F_A^S$) form factors because, unlike the CCQE channel, we now have a simultaneous contribution of these. Figure \ref{fig:6} (top) shows the $\frac{d\sigma}{dQ^2}$ differential cross section for the $\nu_\mu + CH_2$ reaction and the  parametrization of form factors reported in Table \ref{tab:param} (taken from \cite{leitner_neutrino_2005}). It can be observed a very similar behavior of all parametrizations and an underestimation of the experimental data for  $Q < 0.7 \ GeV^2$.

\begin{table}[b]
\caption{\label{tab:param}%
Parameterization of form factors. 
}
\begin{ruledtabular}
\begin{tabular}{rccc}
Parameters & FIT I \cite{garvey_determination_1993} & FIT II \cite{garvey_determination_1993} & FIT III\\
\colrule
$\Delta s$ & $-0.21 \pm 0.10$ & $-0.15 \pm 0.07$ & $0$\\
$F_1^S(0)$ & $0.53 \pm 0.70$ & $0$ & $0$\\
$F_2^S(0)$ & $-0.40 \pm 0.72$ & $0$ & $0$\\ 
$M_A \ (GeV)$ & $1.012 \pm 0.032$ & $1.049 \pm 0.019$ & $1.00$ \\ 
\end{tabular}
\end{ruledtabular}
\end{table}

In figure \ref{fig:6} (bottom), we have the same observable as figure \ref{fig:6} (top), this time using the axial mass, $M_A = 1.35 \ GeV$. It is possible to observe a better reproduction of the experimental data; this demonstrates the need to adopt values of $M_A$ for neutrino-nucleus reactions different from those obtained for neutrino-nucleon reactions. Similar results were obtained by other models, from the adjustment to experimental data. For example, the NUANCE code uses $M_A = 1.35 \ GeV$ to reproduce the MiniBooNE experiment data \cite{aguilar-arevalo_measurement_2015} \cite{the_miniboone_collaboration_measurement_2010}.

\begin{figure}[hbt!]
	\centering
	\includegraphics[scale=0.4]{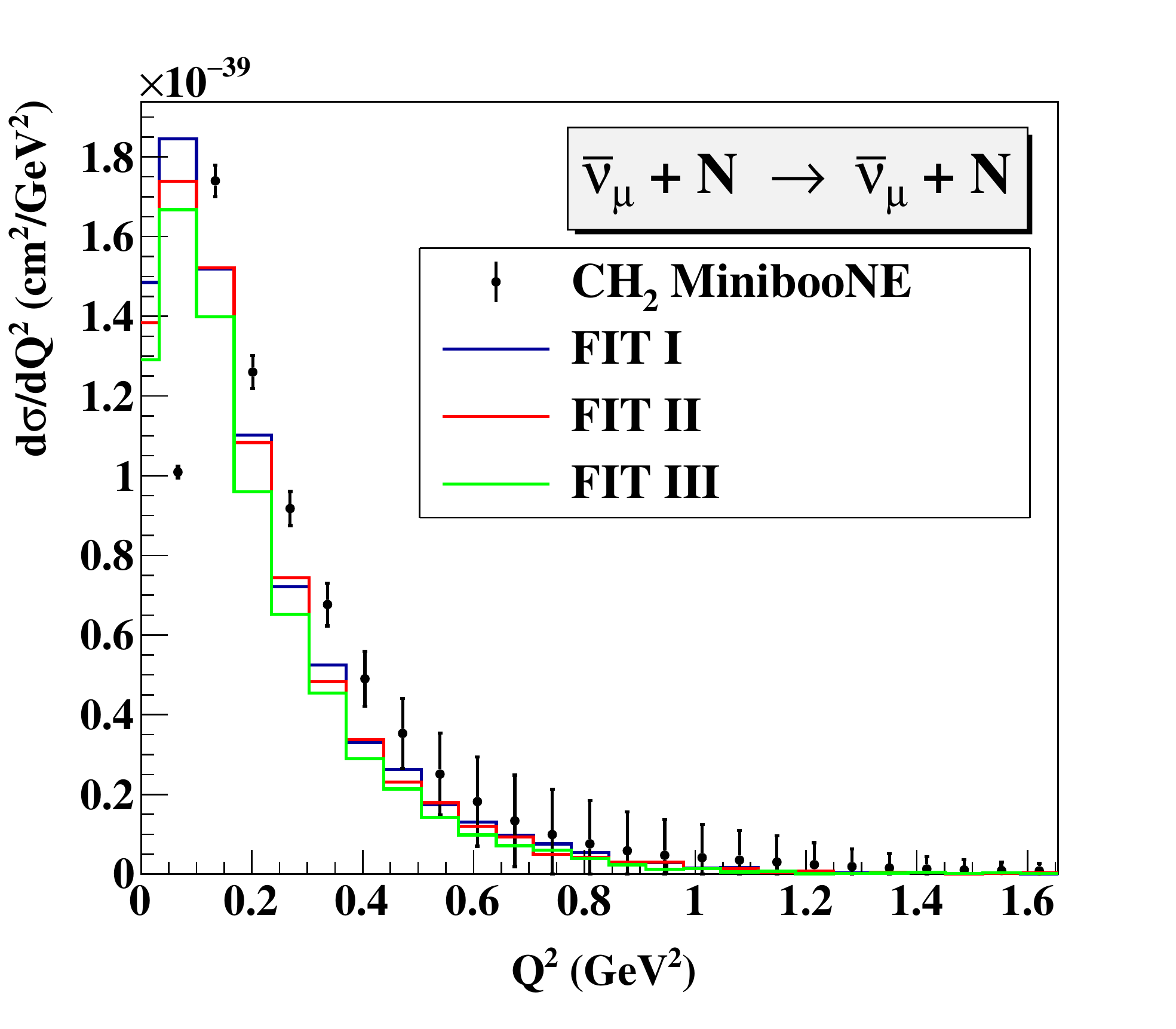}
	\includegraphics[scale=0.4]{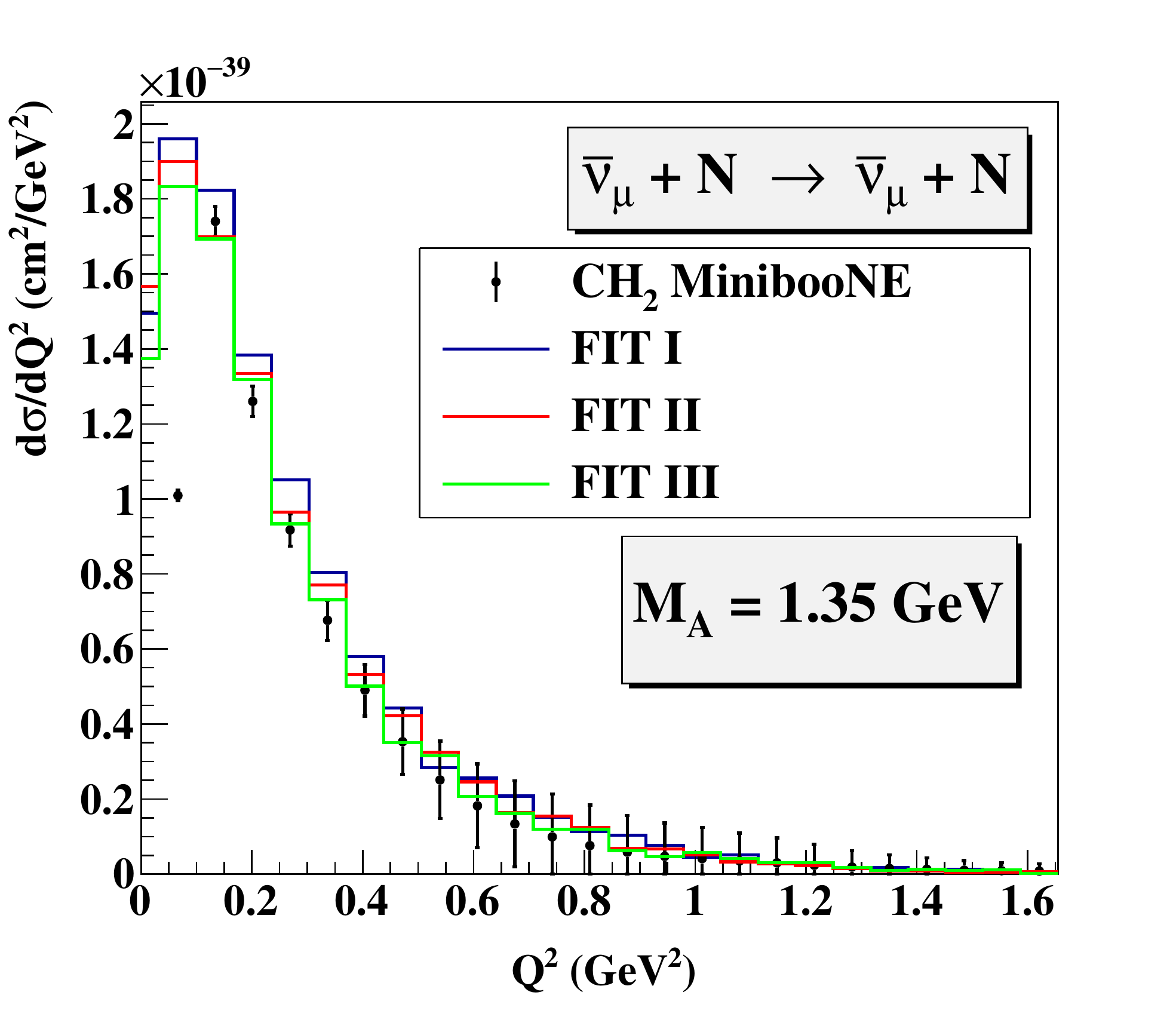}
	\caption{$d\sigma /dQ^2$ cross section for the NCE channel for the reaction $\bar{\nu}_\mu + CH_2$, with parametrizations from table \ref{tab:param}. Top: Parameter $M_A$ from table \ref{tab:param}; Bottom: Parameter $M_A = 1.35 \ GeV$. Experimental data extracted from \cite{aguilar-arevalo_measurement_2015}.} 
	\label{fig:7}
\end{figure}

\subsection{Neutral current production channel of $\pi^0$}

\begin{figure}[hbt!]
	\centering
	\includegraphics[scale=0.4]{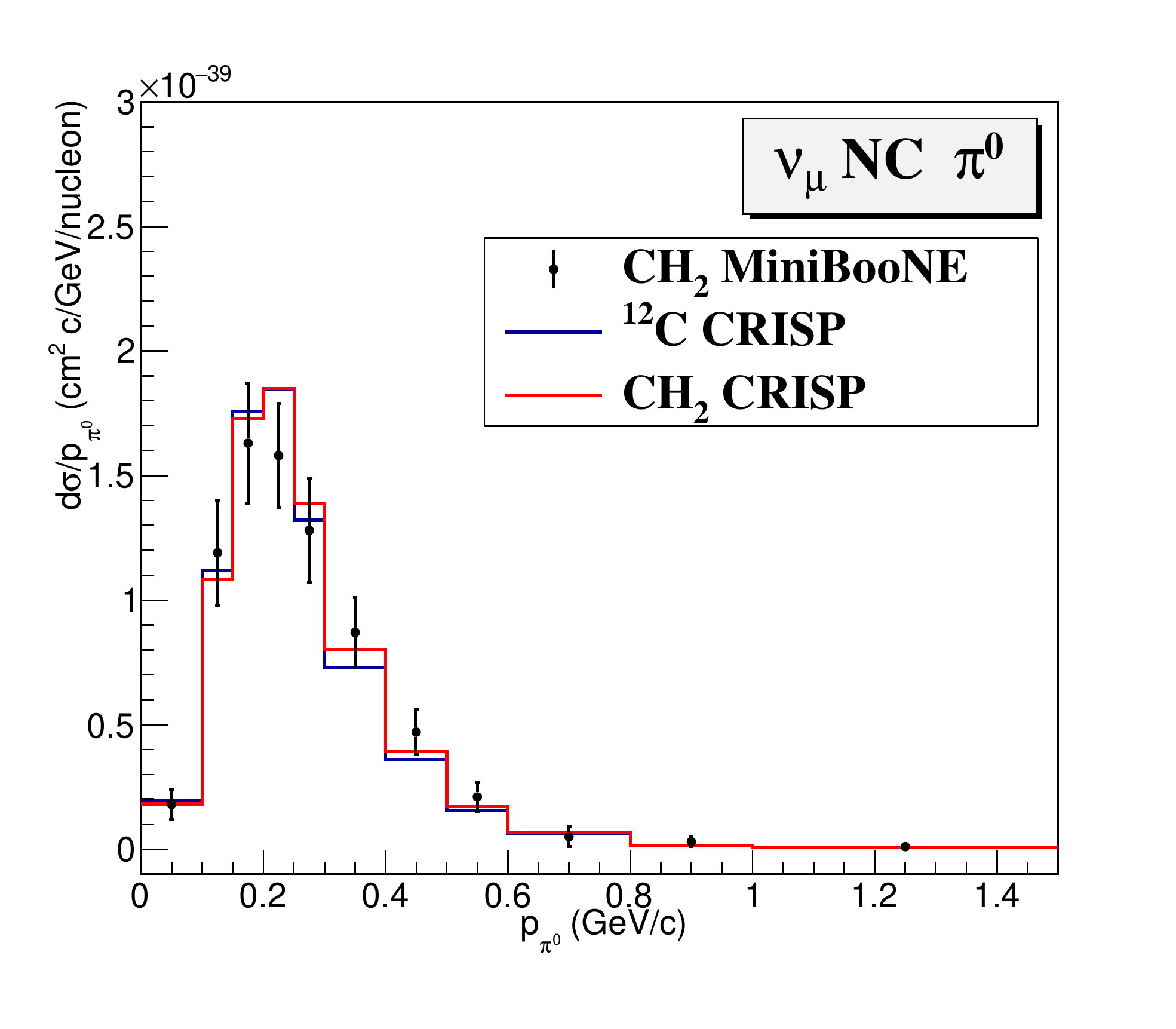}
	\includegraphics[scale=0.4]{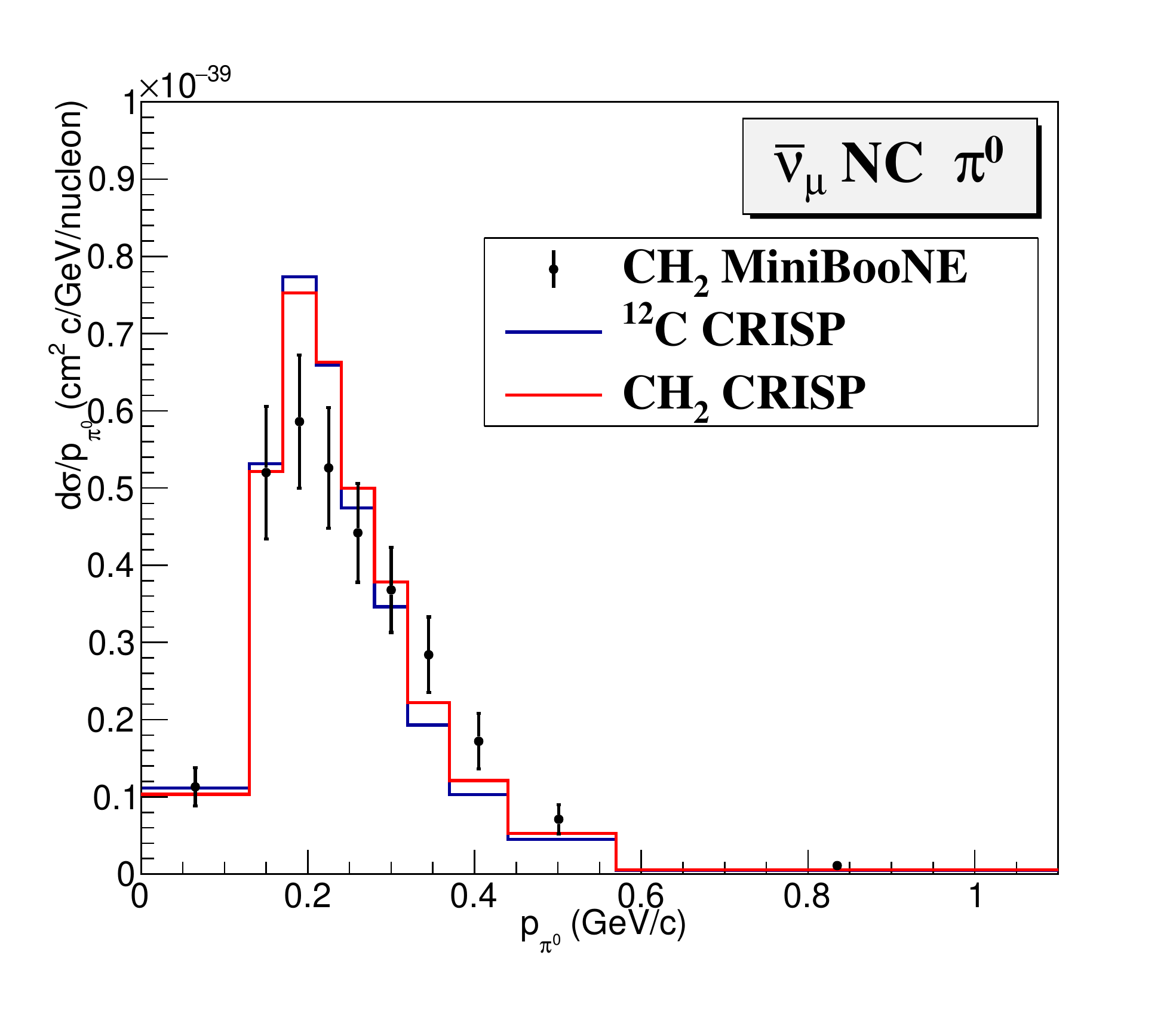}
	\caption{Linear momentum distribution of emitted NC $\pi^0$ during the intranuclear cascade. Experimental data extracted from \cite{aguilar-arevalo_measurement_2015}.}
	\label{fig:71}
\end{figure}

The production of neutral current $\pi^0$ is measured when there is only one emitted meson ($\pi^0$), and there is no muon emission. No restriction is applied to the emission of nucleons \cite{aguilar-arevalo_measurement_2015}. It can be seen how the CRISP model manages to correctly reproduce the shape of the momentum cross section of the emitted pions (\ref{fig:71}). CRISP underestimates the experimental cross section for $p_{\pi^0} > 0.25 \ GeV/c$; it may be associated with the fact that the coherent neutrino-nucleus interaction is not included in our computational model. For the momentum integrated cross section, the $\nu + CH_2$ calculated cross section is $\sigma^{NC\pi^0}_{CRISP} = 4.76 \times 10^{-40} \ cm^2$/nucleon, which is in concordance with the experimental result of $\sigma^{NC\pi^0}_{exp} = 4.76 \times 10^{-40} \ cm^2$/nucleon. Similarly, for $\bar{\nu}_\mu+ CH_2$ interaction, we have that $\sigma^{NC\pi^0}_{CRISP} = 1.49 \times 10^{-40} \ cm^2$/nucleon vs $\sigma^{NC\pi^0}_{exp} = 1.48 \times 10^{-40} \ cm^2$/nucleon. Although the CRISP does not precisely reproduce this differential cross section, it can be seen that the integral value is correctly calculated.

\begin{figure}[hbt!]
	\centering
	\includegraphics[scale=0.4]{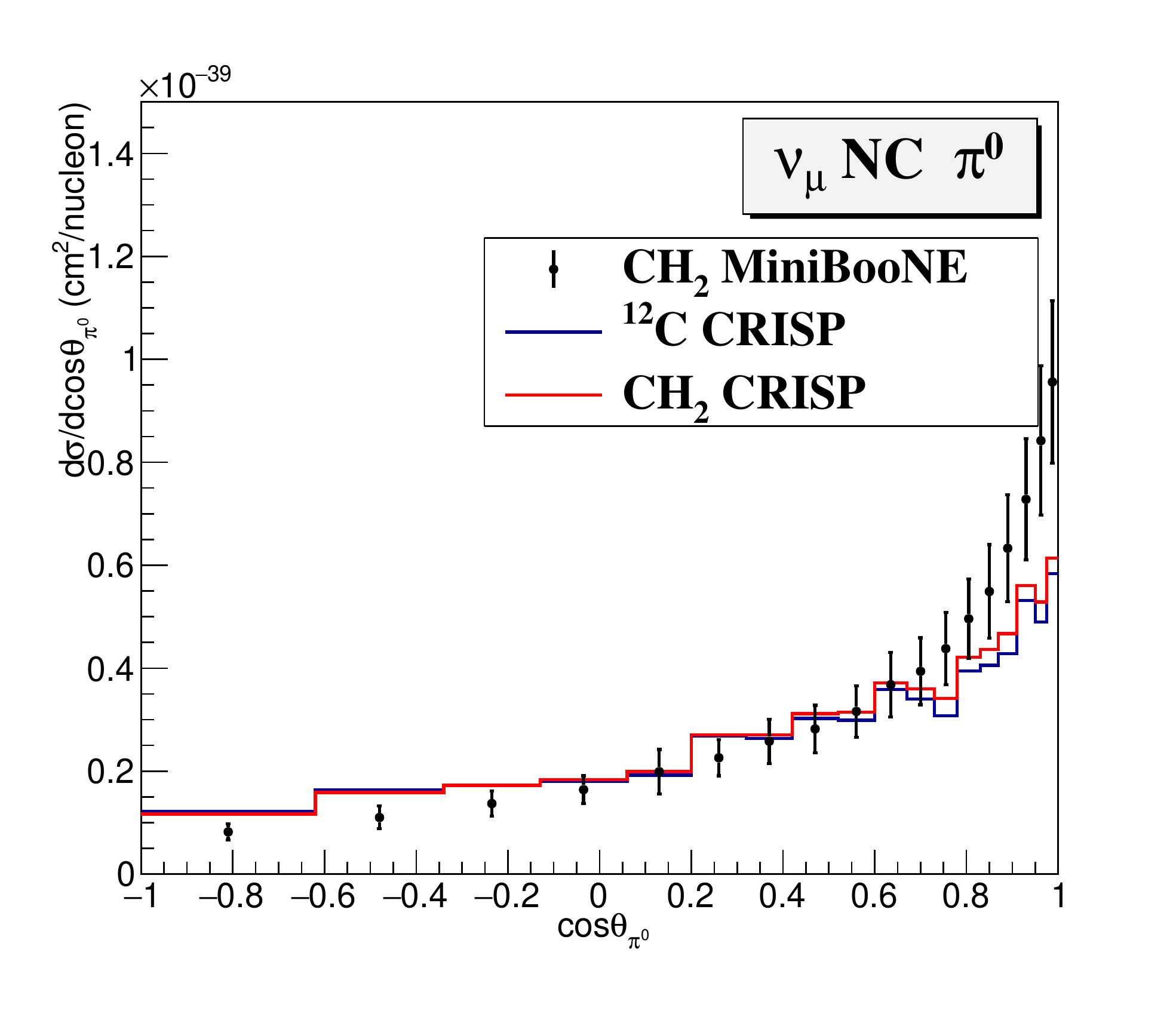}
	\includegraphics[scale=0.4]{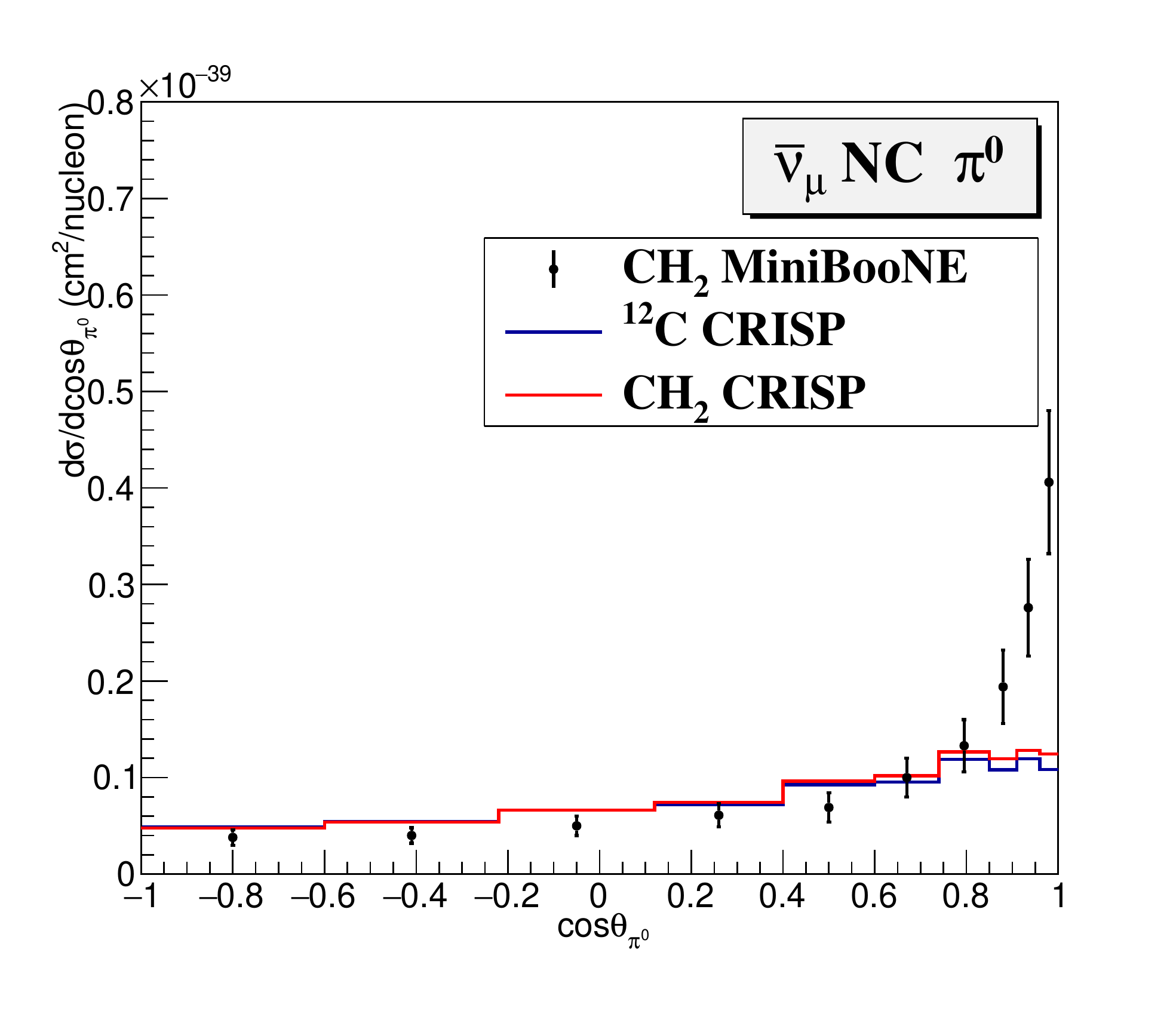}
    \caption{Angular distribution of emitted NC $\pi^0$ during the intranuclear cascade. The emission angle is taken regarding the incident neutrino. Experimental data extracted from \cite{aguilar-arevalo_measurement_2015}.}
	\label{fig:72}
\end{figure}

The super production of pions in the interval $0.15<p_\pi^0 < 0.25 \ GeV/c$ may be associated with the fact that the pions are not being absorbed or exchange charge in an incorrect way in the intranuclear cascade. This over-production is reflected at lower pion emission angles, figure \ref{fig:72}, where the CRISP angular distribution is more homogenized than the experimental data.

\begin{figure}[hbt!]
	\centering
	\includegraphics[scale=0.4]{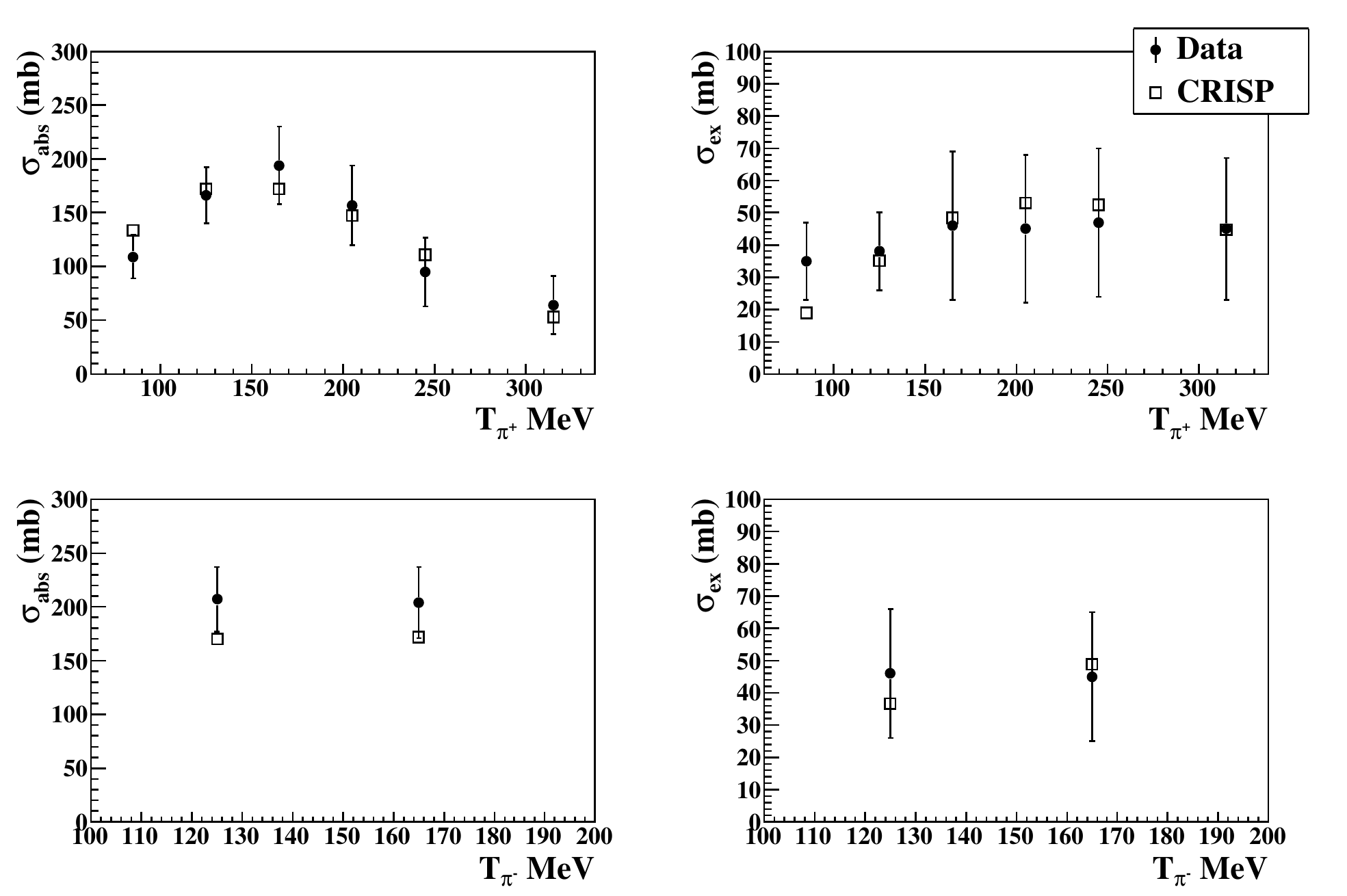}
	\caption{Absorption cross section $\sigma_{abs}$ and charge exchange $\sigma_{ex}$ for $\pi^\pm$ in $^{12}C$. The absorption cross section is measured when pions are not emitted during the intranuclear cascade. The charge exchange cross section is calculated when only one pion is emitted with charge different from the incident pion. Experimental data extracted from \cite{ashery_true_1981}.}
	\label{fig:722}
\end{figure}

To check how pions interact inside the nucleus, we present a CRISP calculation of the absorption and charge exchange cross section for the $\pi^\pm + ^{12}C$ reactions (figure \ref{fig:722}). The pion absorption in CRISP occurs in two ways: the direct absorption by a quasi-deuterium pair ($\pi+(np) = NN$) or the absorption by a nucleon forming a baryonic resonance and the subsequent interaction of this resonance with a nucleon ($\pi + N = \Delta$ and $\Delta + N = N + N$). The isospin symmetry relates the $\pi^+$, $\pi^0$, and $\pi^-$ absorption channels, and therefore, since we have proper absorption of $\pi^+$ and $\pi^-$ é, we expect a good absorption of $\pi^0$.

The following reaction is not implemented in the CRISP yet: the direct pion absorption by two protons or two neutrons ($\pi NN \rightarrow NN$). With the direct pion absorption cross section by a quasi-deuterium pair, which have already been implemented, it can be obtained all the pion plus two nucleon interactions \cite{engel_pion-nucleus_1994}:

\begin{eqnarray}
    \frac{\sigma_{\pi^+(nn)\rightarrow np}}{\sigma_{\pi^+(np)\rightarrow pp}} &= 0.083, \ \  \frac{\sigma_{\pi^0(np)\rightarrow np}}{\sigma_{\pi^+(np)\rightarrow pp}} &= 0.44, \nonumber \\
    \frac{\sigma_{\pi^0(nn)\rightarrow nn}}{\sigma_{\pi^+(np)\rightarrow pp}} &= 0.14, \ \  \frac{\sigma_{\pi^0(pp)\rightarrow pp}}{\sigma_{\pi^+(np)\rightarrow pp}} &= 0.14,  \\
    \frac{\sigma_{\pi^-(pp)\rightarrow np}}{\sigma_{\pi^+(np)\rightarrow pp}} &= 0.083,  \ \  \frac{\sigma_{\pi^-(np)\rightarrow nn}}{\sigma_{\pi^+(np)\rightarrow pp}} &= 1. \nonumber
\end{eqnarray}

The $\pi^{\pm}+ nn$ and $\pi^{\pm}+ pp$ absorption channels have a contribution of 0.083 times the $\pi^{\pm}+ np$ process cross section, which would not represent a significant variation of the results obtained in figure \ref{fig:722}. On the other hand, the net contribution of the $\pi^0$ absorption by a neutron-neutron or proton-proton pair is 0.28, and it may have a notable influence on the $NC1\pi^0$ channel. In this way, a decrease in the $\pi^0$ emission cross section is expected after implementing that new channels. Since the angular $\pi^0$ distribution must decrease, it will not be possible an improvement in the $0.2 < \cos \theta_{\pi^0} < 1$ region, in which the CRISP calculation is already less than the experimental data (figure \ref{fig:72}). It suggests that there is another channel not considered in the CRISP. In effect, we are not considered the coherent neutrino-nucleon reaction channel in our simulations.

The coherent $NC1\pi^0$ neutrino-nucleus channel produces a lepton and a $\pi^0$ with no transferred energy to the target nucleus  \cite{rein_coherent_1983}. In simulations with the NUANCE event generator, other authors  \cite{aguilar-arevalo_measurement_2015}  showed that the coherent neutrino (antineutrino)-$^{12}C$ channel represents the $17\%$ ($38\%$) of the $NC1\pi^0$ cross section. In that work, they demonstrated that only with the coherent reaction's inclusion will it be possible to calculate the angular $NC1\pi^0$ distribution correctly.

\subsection{Charged current production channel of $\pi^+$}

The production of charged current positive pion ($CC1\pi^+$) is measured when there is only one emitted pion ($\pi^0$) and one muon \cite{aguilar-arevalo_measurement_2011}. The most contributing channel to this is the neutrino (antineutrino) resonance formation and the decaying of that resonance:
\begin{equation}
    \begin{tikzcd}[row sep = tiny, column sep = tiny]
        \nu_\mu (\bar{\nu}_\mu) + N \rar & \mu^- (\mu^+) + N^* \arrow[d, shift left = 5ex] \\
        & \hspace{10ex} N^* \rar & N + \pi.
    \end{tikzcd}
    \label{eq.CCRes3}
\end{equation}

The  CCQE interaction can have a contribution to the $CC1\pi^+$ channel:
\begin{equation}
    \begin{tikzcd}[row sep = tiny, column sep = tiny]
       \nu_\mu (\bar{\nu}_\mu) + n(p) \rar &  \mu^- (\mu^+) +  p(n) \arrow[d, start anchor={[xshift=2ex, yshift=1ex]},
end anchor={[xshift=-3.0ex,yshift=0ex]}] \\
       &  \hspace{-4 ex} p(n) + N \to & \hspace{-4 ex} N + N^* \arrow[d, start anchor={[xshift=-0.5ex, yshift=0.5ex]},
end anchor={[xshift=-3.5ex,yshift=0ex]}] \\
       & & \hspace{-8 ex} N^* \to & \hspace{-6 ex} N + \pi.
    \end{tikzcd}
    \label{eq.CCRes4}
\end{equation}
\begin{figure}[hbt!]
			\centering
			\includegraphics[scale=0.4]{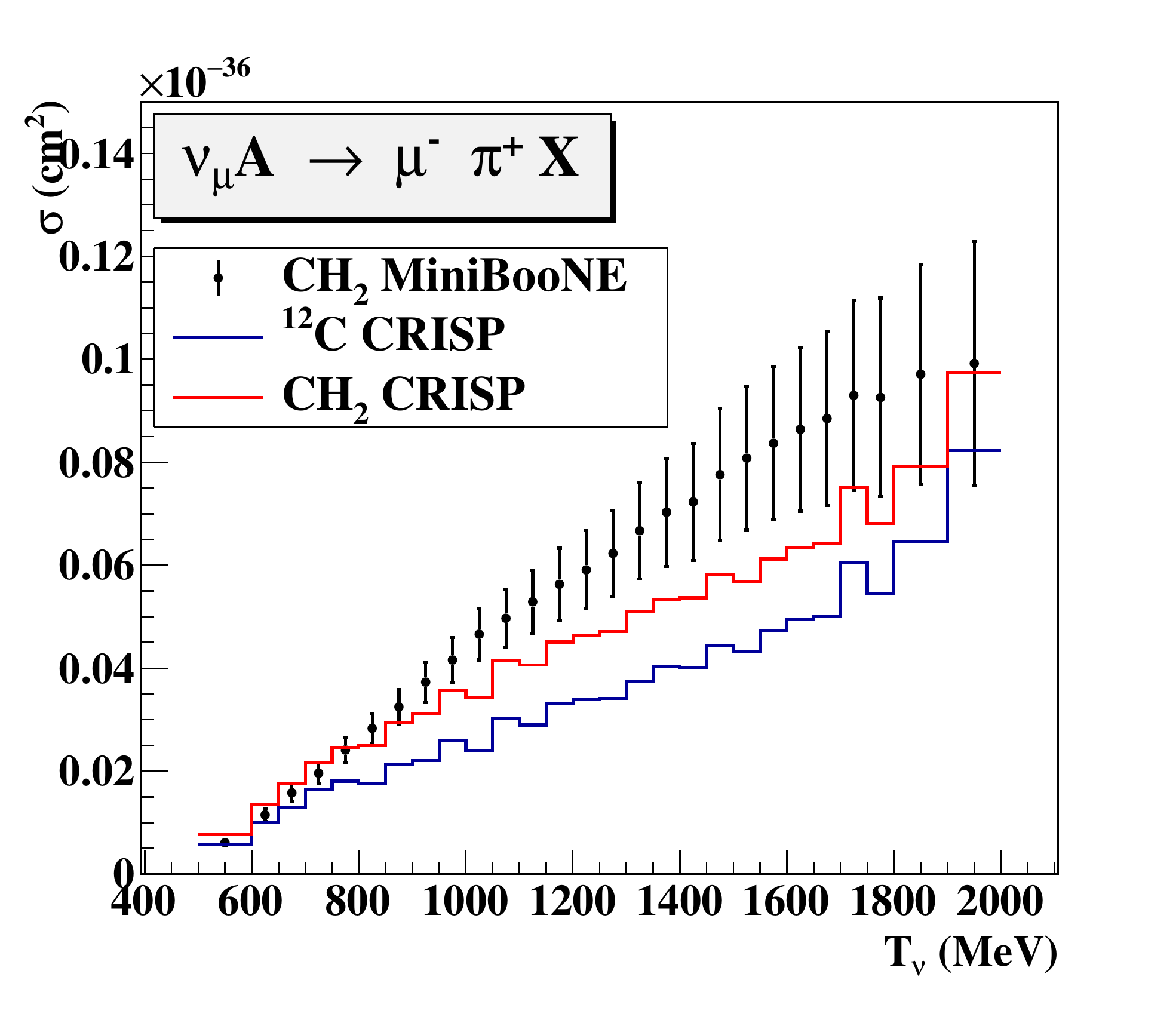}	
			\includegraphics[scale=0.4]{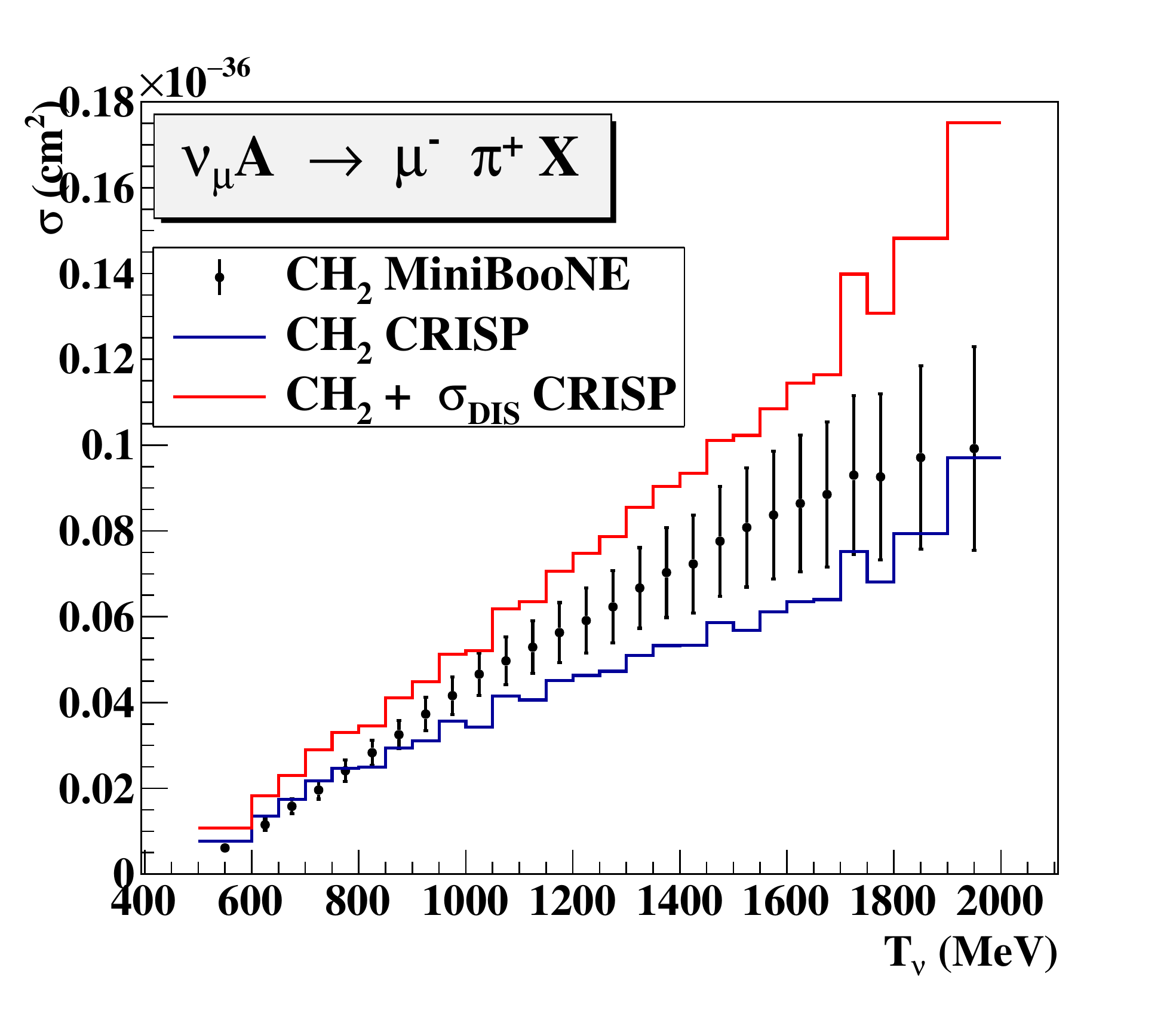}	
			\caption{CC $\pi^+$ emission cross section for the reactions $\nu_\mu + ^{12}C$ and $\nu_\mu + CH_2$ (top). Maximum and minimum contribution of the DIS channel to the $\nu_\mu + CH_2$ interaction (bottom). Experimental data extracted from \cite{aguilar-arevalo_measurement_2011}.}
			\label{fig:81}
\end{figure}
The CRISP model calculated cross section presents a sub estimation of the experimental data (figure \ref{fig:81}).  A potential cause of this is that we do not consider the deep inelastic scattering channel (DIS). DIS has two steps; the first one is the neutrino-quark interaction, and the second one is the hadronization phase. With the first step implementation, we may get a cross section formula and compute this channel's contribution to our results. The hadronization phase consists of the determination of the final state particles. In the case of one $\pi^+$ and one nucleon, the cross section must increase, and therefore a better calculated-experimental match will be obtained.

Figure \ref{fig:81} (bottom) shows the $CC1\pi^+$ cross section considering the maximum possible DIS contribution: the sum of previously $CC1\pi^+$ and DIS cross section.  The blue line represents the $CC1\pi^+$  without the DIS channel, and the red line represents the $CC1\pi^+$  cross section if all DIS processes produce only a $\pi^+$. Any contribution of DIS channel to the $CC1\pi^+$  channel significantly improved CRISP compared with experimental data.

\begin{figure}[hbt!]
			\centering			
			\includegraphics[scale=0.4]{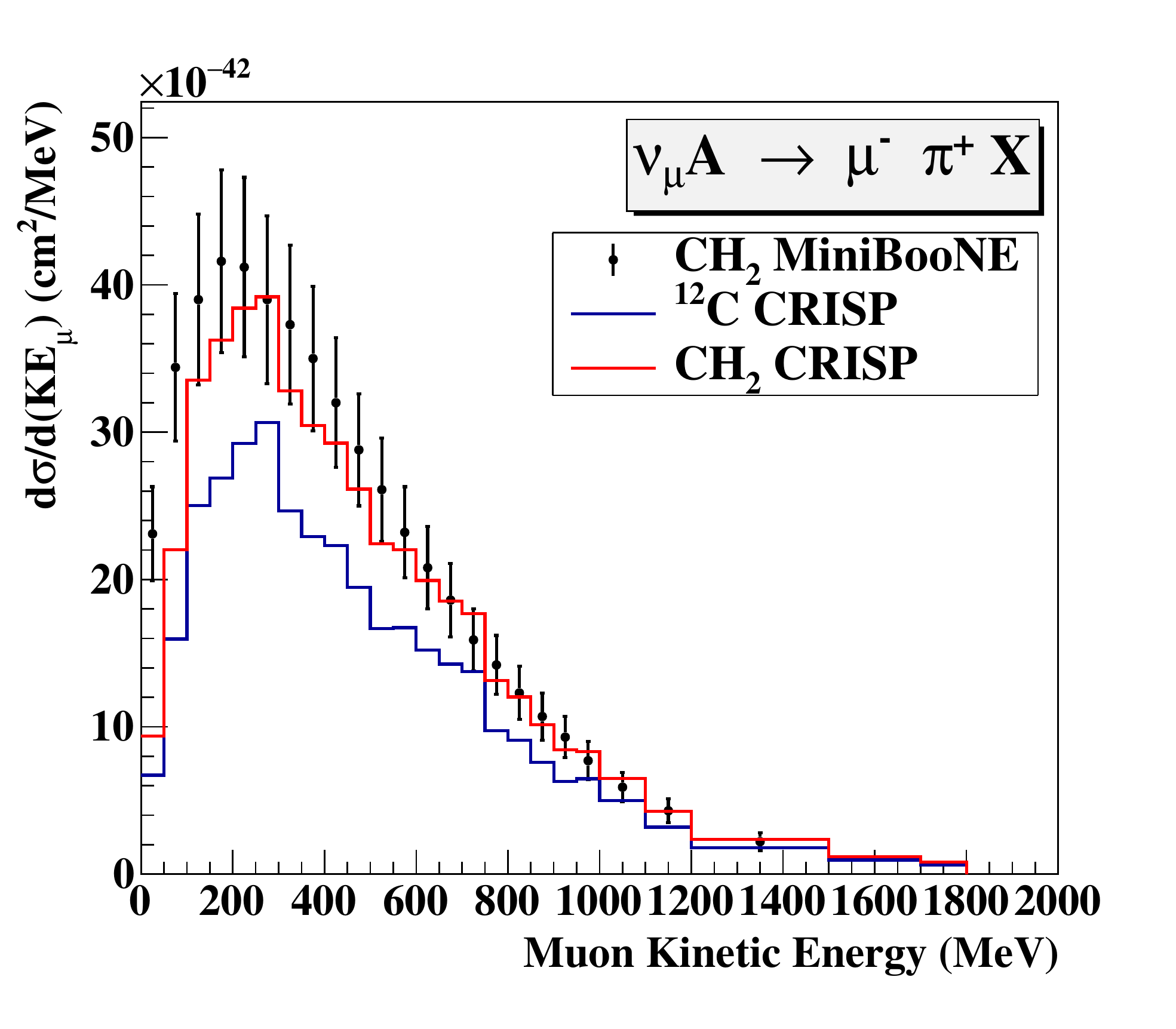}			\includegraphics[scale=0.4]{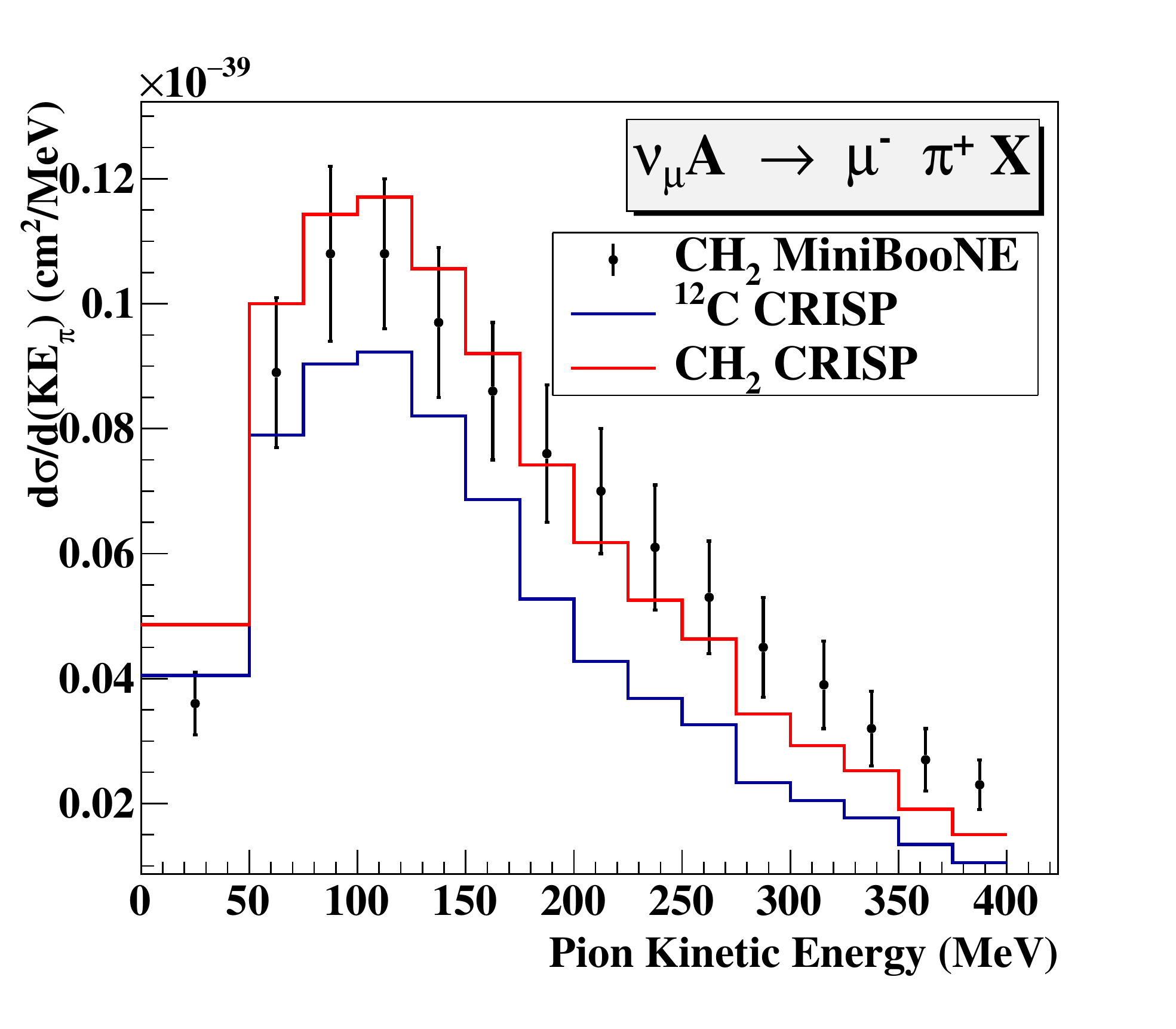}
			\caption{Kinetic energy distributions of emitted CC $\mu^-$ (top) and $\pi^+$ (bottom) for the reactions $\nu_\mu + ^{12}C$  and $\nu_\mu + CH_2$. Experimental data extracted from \cite{aguilar-arevalo_measurement_2011}.}
			\label{fig:10}
\end{figure}

CRISP offers a good reproduction of the $CC1\pi^+$ emitted $\mu^-$ and $\pi+$ kinetic energy distributions, despite the small data underestimation obtained (figure \ref{fig:10}). If we take the energy-dependent cross section (figure \ref{fig:81}), we may expect a more significant underestimation of the experimental data. It does not happen because of the incident neutrino (antineutrino) flux energy distribution. The most significant difference in data-calculation is obtained to higher neutrino energy, where the incident neutrino number is the lowest and, therefore, less its contribution to the cross section. 

\subsection{CRISP application}

\begin{figure}[hbt!]
			\centering			
			\includegraphics[scale=0.3]{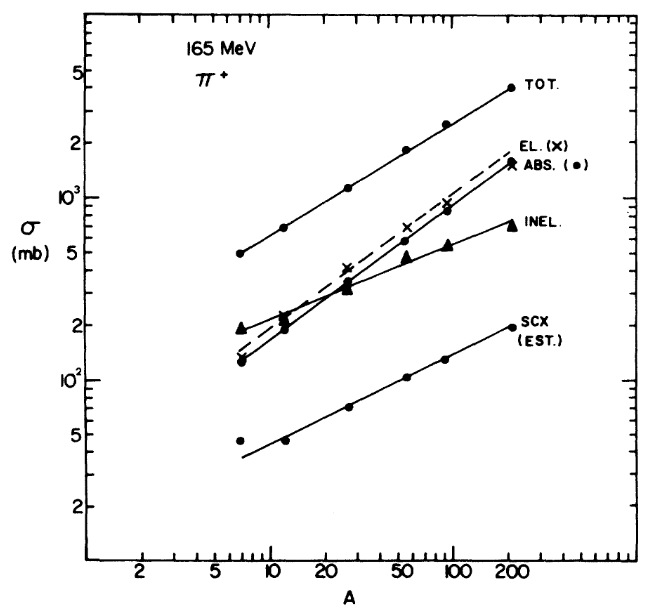}
			\caption{Decomposition of the total $\pi^+$-nucleus cross section at 165 MeV. Figure extracted from \cite{ashery_true_1981}.}
			\label{fig:pion_nucleo}
\end{figure}

We start this section with an analysis of the pion-nucleus reaction. In figure \ref{fig:pion_nucleo} is shown the experimental nuclear mass number dependence of the exclusive cross section for the $\pi^+$-nucleus reaction. If we emphasize the inelastic and absorption channels, we can see a notable difference in the cross section slope. The inelastic channel occurs, in most cases, when the incident $\pi^+$ scatters by pion-nucleon collisions inside the target nucleus in such a way that it leaves the nucleus. On the other hand, that $\pi^+$ cannot be absorbed through pion-nucleon reactions; it has to be absorbed by at least a couple of nucleons, hence the need to introduce a pion-nucleon-nucleon absorption mechanism to describe the pion-nucleus reaction. In this way, as the mass number of the target nucleus increases, the number of possible pairs of nucleons that can absorb the incident $\pi^+$ increase more than the number of nucleons that can scatter it, and therefore, the absorption channel slope is greater than the inelastic channel slope.

In general, we can establish that the cross section dependence with the target mass number offers relevant information about the primary interaction nature, specifically, if the incident particle interacts with target nucleons independently or with more than one simultaneously. We will apply this methodology and determine if the neutrino-nucleon interaction is sufficient to describe the neutrino nucleus reactions under our intranuclear cascade formalism.

\begin{figure}[hbt!]
			\centering			
			\includegraphics[scale=0.4]{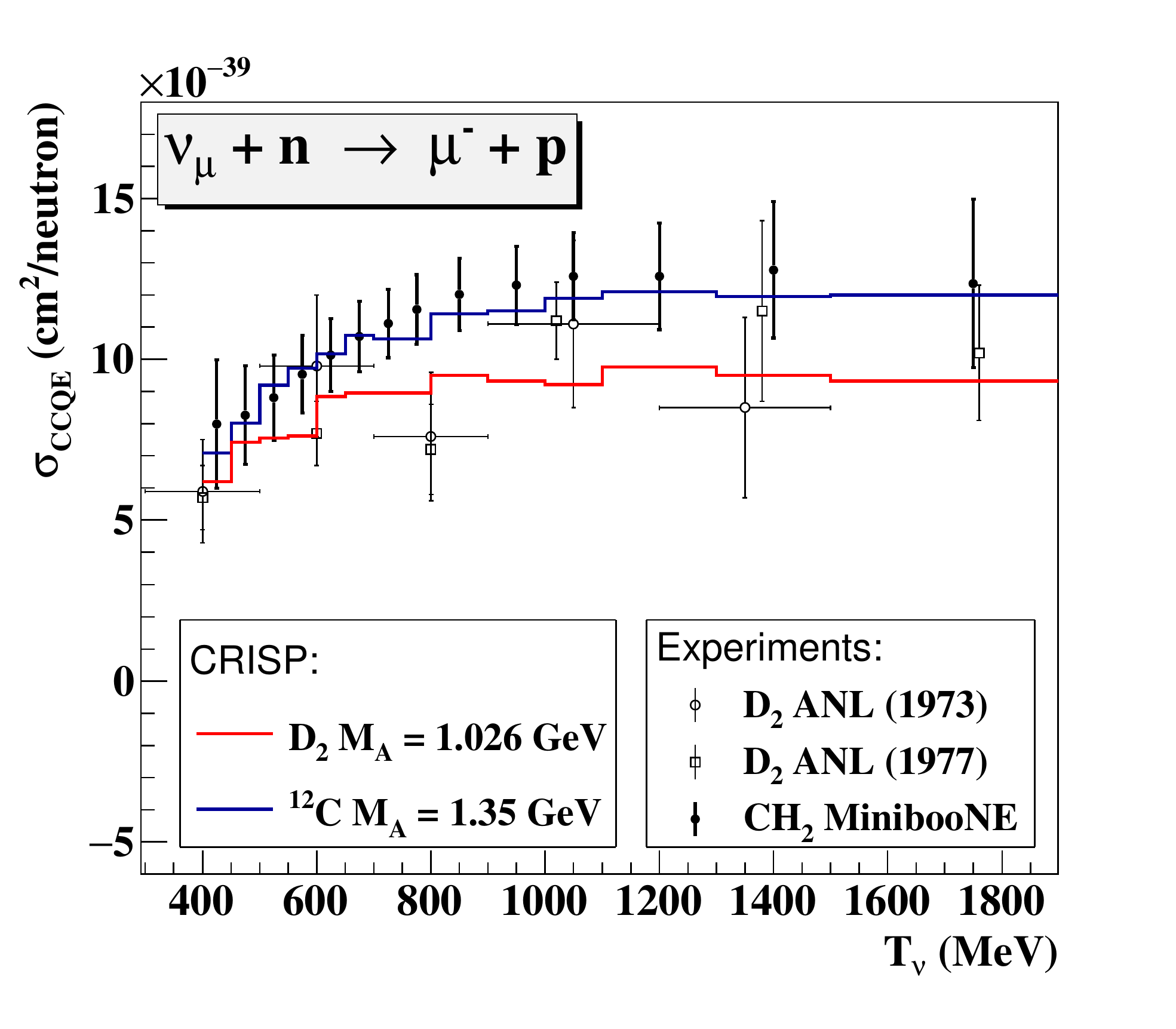}			\includegraphics[scale=0.4]{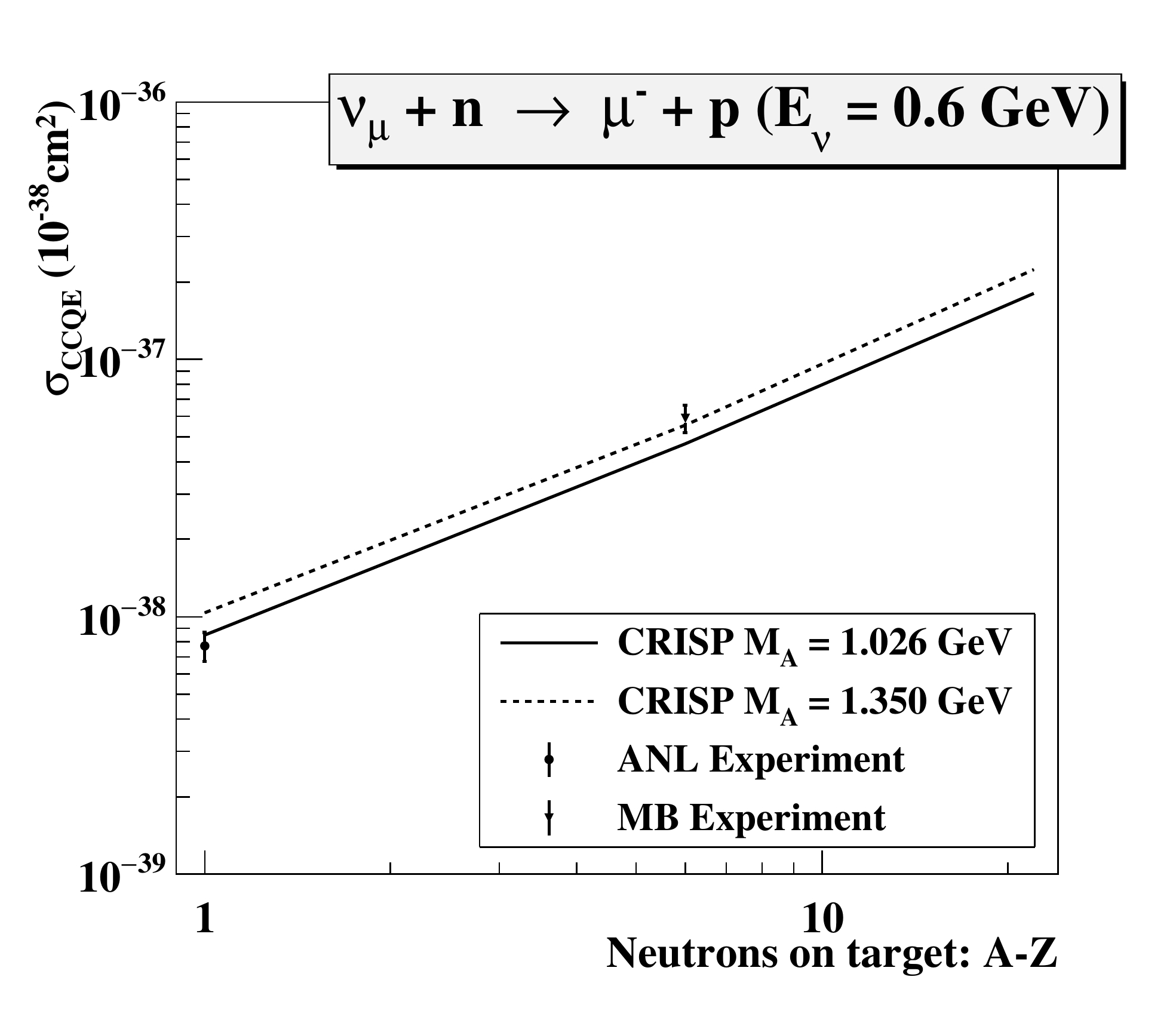}
			\caption{Top: Experimental (points) and calculated (lines) CCQE neutrino-neutron cross section on deuterium ($D_2$) and $^{12}C$. Bottom: Dependence of the experimental (points) and calculated (lines) cross section with the target mass number.}
			\label{fig:crisp_exp}
\end{figure}

Figure \ref{fig:crisp_exp} (top) shows the CCQE neutrino-neutron cross section for incident muon neutrinos on deuterium and $^{12}C$. The experimental cross section on $^{12}C$ is higher than on deuterium but, that should not happen since, on $^{12}C$, the Pauli blocking mechanism is more effective than on deuterium. For this reason, it is necessary to use a different axial mass parameter $M_A$ in the $^{12}C$ reaction ($M_A = 1.35 \ GeV$) than in the neutrino nucleon interaction ($M_A = 1.026 \ GeV$ \cite{bernard_axial_2002}).  In the IV B section, we showed that $M_A = 1.012 \ GeV$ does not reproduce the neutrino-$^{12}C$ experimental results. 

Figure \ref{fig:crisp_exp} (bottom) represents the neutrino nucleus cross section as a function of the number of neutrons on the target nucleus. We selected $E_\nu = 0.6 \ GeV$ because, for that energy, the experimental data reasonably agree with the theoretical predictions (figure \ref{fig:crisp_exp} at top). It can be observed that with the variation of $M_A$, the linear behavior of $\sigma_{CCQE}$ does not have a change in the slope, and therefore we can conclude that under the adopted intranuclear cascade model, it will not be possible to reproduce the deuterium and $^{12}C$ data with the same $M_A$ parameter.

\begin{figure}[hbt!]
			\centering
			\includegraphics[scale=0.20]{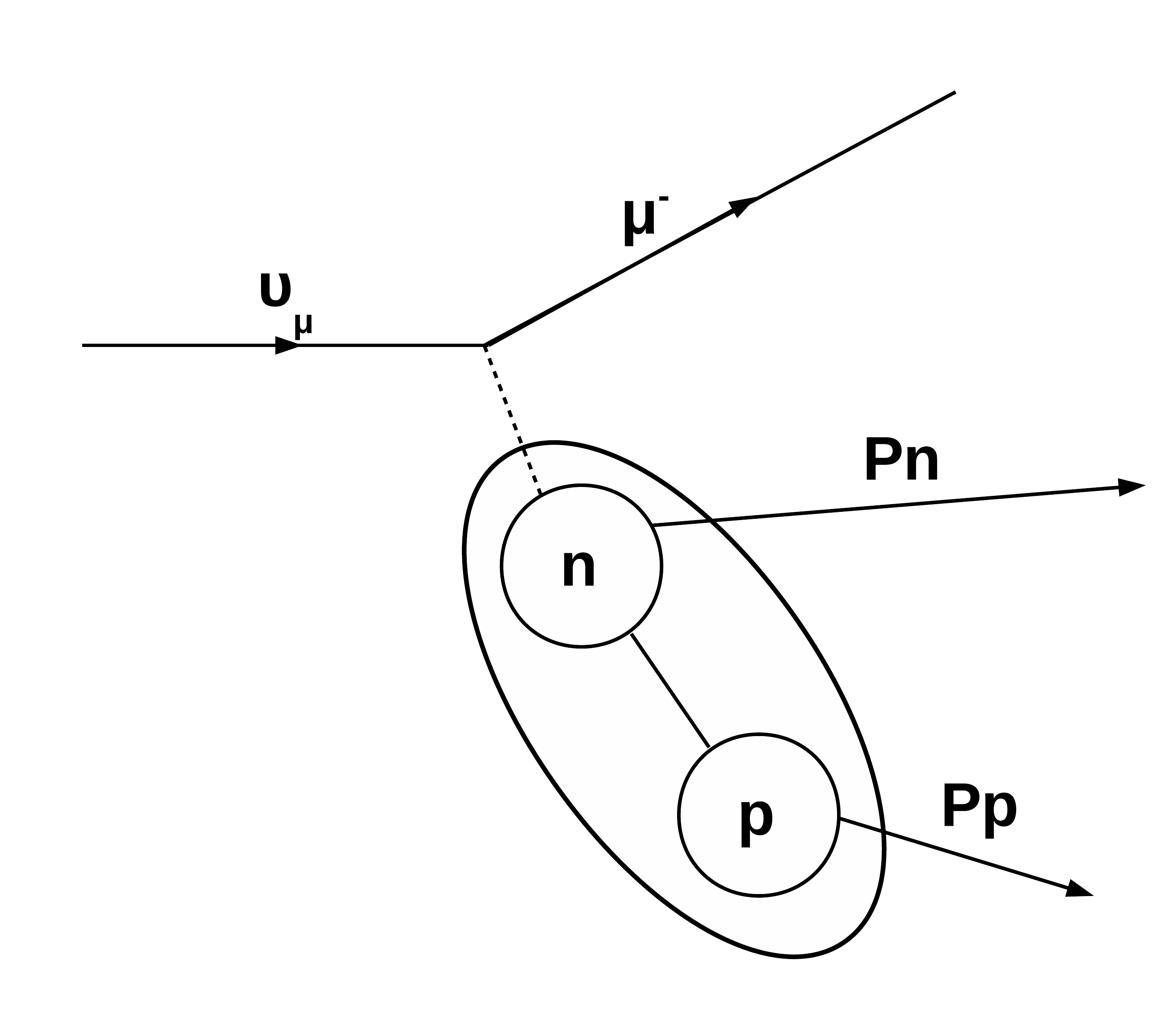}
			\includegraphics[scale=0.4]{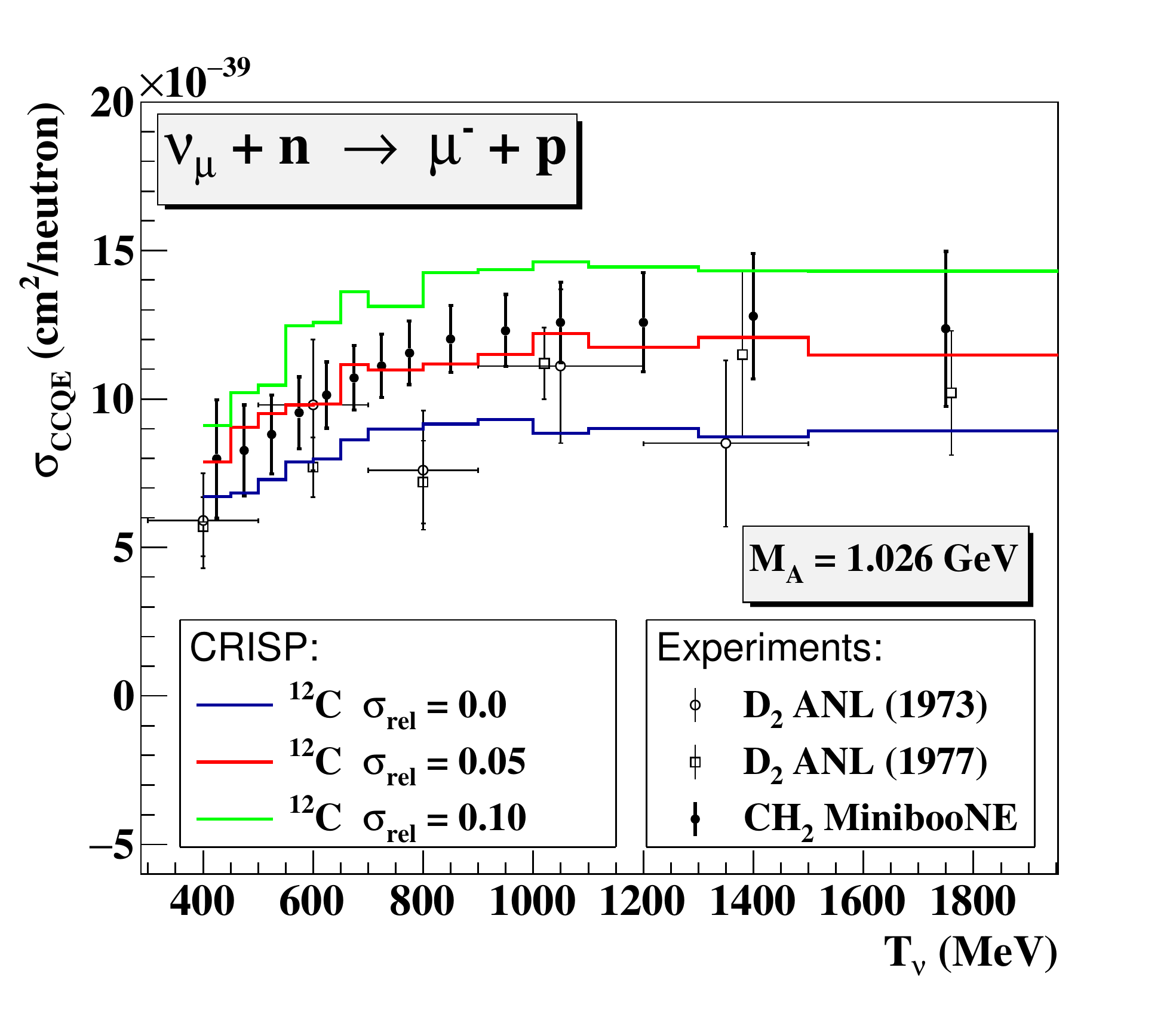}			
			\caption{Top: Experimental (points) and calculated (lines) CCQE neutrino-neutron cross section on deuterium ($D_2$) and $^{12}C$. Bottom: Dependence of the experimental (points) and calculated (lines) cross section with the target mass number.}
			\label{fig:d2_model}
\end{figure}

To get an increment of the $\sigma_{CCQE}$ slope, we suppose that the incident neutrino can interact with more than one nucleon simultaneously. The most simple model we can consider is the neutrino-nucleon-nucleon interaction. Physically, the reaction of the neutrino with a di-nucleon system occurs when the neutrino interacts with a nucleon while the two nucleons are interacting with each other.  We have represented a simplified scheme of this reaction in figure \ref{fig:d2_model} (top).

To introduce the neutrino-nucleon-nucleon collisions in the CRISP code, the following CCQE cross section was adopted:
\begin{equation}
    \sigma = \sigma_{\nu N} + \sigma_{\nu NN},
    \label{sigma_rel_1}
\end{equation}
where $\sigma_{\nu N}$ is the cross section when the neutrino interacts with a proton of the nucleon-nucleon pair and $\sigma_{\nu NN}$ is the cross section when the neutrino reacts with the two nucleons according to the scheme explained in figure \ref{fig:d2_model} (top).

As $M_A$ was calculated using the experimental neutrino-deuterium data, we can consider as a first approximation that $\sigma$ is calculated using equation \ref{CCQE_sigma_dif}. Thus, with the objective of estimating $\sigma_{\nu N}$ and $\sigma_{\nu NN}$, the following relative cross section was defined:

\begin{equation}
    \sigma_{rel} = \frac{\sigma_{\nu NN}}{\sigma}.
    \label{sigma_rel_2}
\end{equation}

Thus, if are given $\sigma$ (figure \ref{CCQE_sigma_dif}) and $\sigma_{rel}$ (put manually), then it is possible to calculate $\sigma_{\nu N}$ and $\sigma_{\nu NN}$ using equations \ref{sigma_rel_1} and \ref{sigma_rel_2}. It would be strictly necessary to consider that $\sigma_{rel}$ depends on the energy of the neutrino, but to obtain a general result, we can assume that we are working with the mean value of $\sigma_{rel}$. Figure \ref{fig:d2_model} shows the calculated $\sigma_{CCQE}$ with the CRISP model for different values of $\sigma_{rel}$, it can be noticed how it is possible to reproduce the experimental data for $^{12}C$ using $M_A = 1.026 \ GeV$ and $\sigma_{rel} = 0.05$.

\begin{figure}[hbt!]
			\centering
			\includegraphics[scale=0.4]{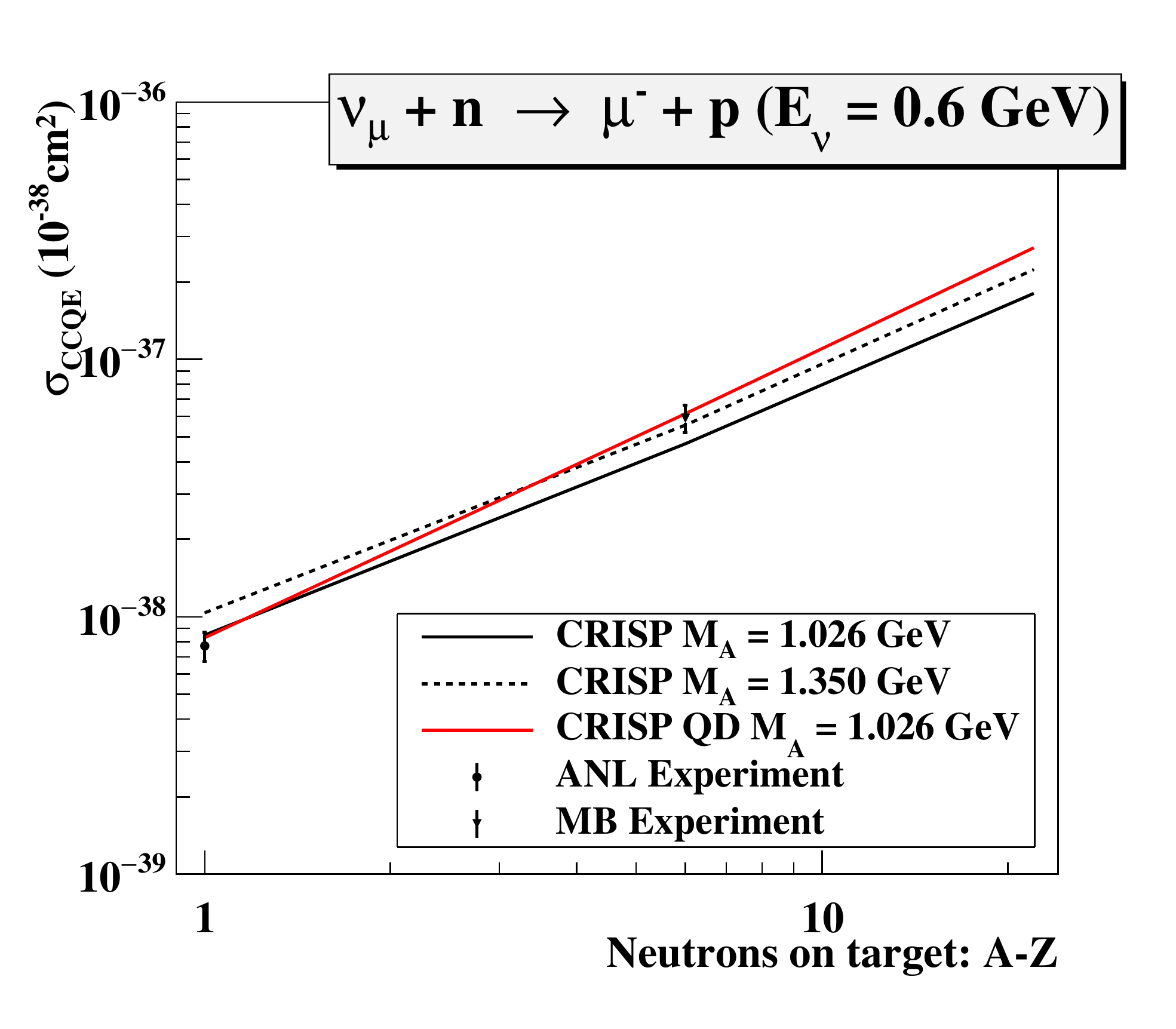}			
			\caption{Dependence of the experimental (points) and calculated (lines) cross section with the target mass number.The black lines are the CRISP calculations under the neutrino-nucleon interactions. The red line represents the CRISP calculations under the neutrino-nucleon and the neutrino-nucleon-nucleon interactions.}
			\label{fig:final}
\end{figure}

\begin{table}[b]
\caption{\label{tab:table1}%
CRISP predictions of the $\sigma_{CCQE}$ for the $\nu_\mu + ^{40}Ar$ reaction. 
}
\begin{ruledtabular}
\begin{tabular}{ccc}
$M_A$ (GeV) & $\sigma \ (10^{-37}$ $cm^2$) & neutrino interactions \\
\colrule
1.026       & 1.802  &      $\nu_\mu + N$       \\
1.35        & 2.237  &        $\nu_\mu + N$        \\ 
1.026  & 2.709  &       $\nu_\mu + N$, $\nu_\mu + NN$         \\ 
\end{tabular}
\end{ruledtabular}
\end{table}

Now it is possible to simultaneously reproduce the experimental neutrino-deuterium and neutrino-$^{12}C$ $\sigma_{CCQE}$, both with the same value of $M_A = 1.026 \ GeV$. Figure \ref{fig:final} shows how with the new model, an increase in the sigma slope line is obtained, just as desired. To determine if there is a significant neutrino-nucleon-nucleon interaction contribution, measurements on heavier nuclei will be necessary. For example, we present the $\nu_\mu + ^{40}Ar$ reaction, which will be studied soon in the DUNE \cite{Dune_page} experiment. Table 1 shows our $\sigma_{CCQE}$ predictions for this reaction, according to the form that the neutrino interacts (with one or two nucleons) and the $M_A$ parameters studied in this work.

\section{Conclusions}

In this work we presented a neutrino-nucleon interaction model integrated with an intranuclear cascade model part of CRISP package. Both charged and neutral currents were calculated while examining the effect of the nuclear medium on the cross section results. Thanks to the multicollisional approach and the strict Pauli blocking verification, it was possible to observe that the fermionic motion of the nucleons as well as the Pauli blocking mechanism have distinctive and appreciable effects both at the momentum transferred distribution and the double differential cross sections. In addition, the nucleon separation energy was observed to have a major effect on the momentum transferred distribution.

The present model does not take into account coherent neutrino-nucleus interaction which prevents it from reproducing the correct $\pi^0$ angular distribution as emitted in the neutral current channels with an underestimation of the cross section for small emission angles. Deep inelastic scattering is also not considered, which leads to an underestimation of the cross section of the charged current production channel of $\pi^+$ in terms of the neutrino kinetic energy. Despite that, the model provides a good reproduction of both muon and pion kinetic energy distributions. Therefore, the model was found to provide excellent agreement with most experimental data, while the underestimations observed can be readily explained by known missing channels.

By paralleling the pion-nucleus reaction, we introduce an $\nu-NN$ interaction model in the intranuclear cascade. In this way, it was possible to simultaneously reproduce the experimental data for $\nu_\mu + d_2$ and $\nu_\mu + ^{12}C$ reactions with the same axial mass value. The inclusion of the $\nu-NN$ interaction had a notable influence on sigma dependence on the target mass number. That can be used in future experiments to determine if this interaction is present in the neutrino-nucleus reaction. As an example calculation, we present our predictions for the $\nu_\mu + Ar$ reaction, which will soon measure in the DUNE \cite{Dune_page} experiment.

\section{Acknowledgments}
Ramon Perez acknowledges the support from CNPq under Grant 169813/2017-7. Airton Deppman is supported by the Project INCT-FNA Proc. No. 464898/2014-5 and by FAPESP Grant 2016/17612-7. Airton Deppman is also partially supported by Conselho Nacional de Desenvolvimento Científico e Tecnológico CNPq (Brazil) under Grant 304244/2018-0.

\bibliography{apssamp}

\end{document}